\documentclass[twocolumn]{aastex63}
\usepackage{inputenc}
\usepackage{comment}
\usepackage{colortbl}

\received{14 Mar 2024}
\revised{2 May 2024}
\accepted{2 May 2024}
\submitjournal{ApJ}

\shorttitle{4FGL Blazar SEDs}
\shortauthors{Kerby \& Falcone}
\graphicspath{{./}{figures/}}
\begin{document}

\title{Testing the Blazar Sequence with Spectra of Recently Discovered Dim Blazars from the \textit{Fermi} Unassociated Catalog}

\author[0000-0003-2633-2196]{Stephen Kerby}
\affiliation{Department of Astronomy and Astrophysics \\
 Pennsylvania State University,
University Park, PA 16802, USA}

\author[0000-0002-5068-7344]{Abraham D. Falcone}
\affiliation{Department of Astronomy and Astrophysics \\
 Pennsylvania State University,
University Park, PA 16802, USA}

\begin{abstract}

Recent works have developed samples of blazars from among the \textit{Fermi}-LAT unassociated sources via machine learning comparisons with known blazar samples. Continued analysis of these new blazars tests the predictions of the blazar sequence and enables more flux-complete samples of blazars as a population. Using \textit{Fermi}, \textit{Swift}, WISE, and archival radio data, we construct broadband spectral energy distributions for 106 recently identified blazars. Drawn from the unassociated 4FGL source sample, This new sample has a lower median flux than the overall sample of gamma-ray blazars.  By measuring the synchrotron peak frequency, we compare our sample of new blazars with known blazars from the 4LAC catalog. We find that the bulk of the new blazars are similar to High-Synchrotron Peak (HSP) BL Lac objects, with a higher median synchrotron peak; the sample has a median $\log (\nu_{syn}/\rm{Hz}) = 15.5$ via \verb|BLaST| peak estimation, compared to $\log (\nu_{syn}/\rm{Hz}) = 14.2$ for the 4LAC BL Lacs. Finally, we conduct synchrotron self-Compton (SSC) leptonic modeling, comparing fitted physical and phenomenological properties to brighter blazars. We find that the new blazars have smaller characteristic Lorentz factors $\gamma_{boost}$ and fitted magnetic fields $B$, in agreement with blazar sequence predictions. The new blazars have slightly higher Compton dominance ratios than expected, which may point to alternative emission models for these dim blazars. Our results extend the predictions of the blazar sequence to a sample of dimmer blazars, confirming the broad predictions of that theory.

\end{abstract}


\section{Introduction}
\label{sec:Intro}

In the unified scheme of active galactic nuclei (AGN) \citep{Blandford1978,Urry1995}, blazars are radio-loud AGN with a jet viewed almost directly on-axis. Electromagnetic emission from blazars is dominated by nonthermal processes producing two broad SED components; a low-energy bump (spanning from radio- to X-rays) attributed to synchrotron emission \citep{Marashi1992,Celotti2008} and a high-energy gamma-ray bump attributed to synchrotron self-Compton, external Compton upscattering, and/or hadronic processes like proton synchrotron and  hadronic cascades \citep{Dermer1993,Mannheim1993,Mucke2001,Mucke2003,Celotti2008,Abdo2010}. Given their extremely broadband emission, blazars can be studied across the electromagnetic spectrum, serving as laboratories for high-energy and relativistic astrophysics in the nearby and distant universe. Furthermore, due to their prolific high-energy emission, blazars numerically dominate catalogs of gamma-ray point sources like the \textit{Fermi} Large Area Telescope (\textit{Fermi}-LAT) source catalog \citep{4FGLorig}.

Blazars are divided into two subcategories based on the equivalent width of optical emission lines, BL Lacs having $\rm{EW}<5 \AA$ and FSRQs having $\rm{EW}>5 \AA$ \citep{Stickel1991}.  Many blazars have high degrees of flux or spectral variability, or periodically enter flaring states; at different times, these changes make them appear alternatively as BL Lacs or FSRQs, blurring the lines between the two categories \citep{Tavecchio1998,Abdo2010b,Bianchin2009}. In general, FSRQs tend to dominate at higher redshifts and have lower synchrotron peak frequencies ($\nu_{syn}$), while BL Lacs are more typically found at low redshift and have higher synchrotron peak frequencies. Because BL Lacs lack optical emission lines, it is generally more difficult to constrain their redshift with spectroscopy, and many BL Lacs have unknown or highly uncertain redshift estimates. The general differences between these two subtypes and the relations between observable properties of blazars are long-running topics of current research involving more detailed study of known blazars and the expansion of catalogs with new discoveries.

The most recent release of the \textit{Fermi}-LAT source catalog, 4FGL-DR3 \citep{4FGLDR3, 4FGLorig}, contains 6658 sources, of which 3676 are associated with blazars. These blazars are further subdivided into BL Lacs and FSRQs, plus numerous ``blazars of uncertain type" (bcu) that are likely BL Lacs and FSRQs that currently defy classification. The 4FGL catalogs represent the most complete list of gamma-ray blazars, a foundation of population studies of jetted AGN.  Blazars make up a plurality of known gamma-ray sources, allowing for numerous population studies starting from high-energy observations. The blazars of the 4FGL catalog are examined in the 4LAC catalogs \citep{4LACDR3, 4LACorig} with spectral fitting, detailed associations, and in-depth classification.

Originating in \cite{Fossati1998} and expanded in subsequent works including \cite{Ghisellini2008} and \cite{Ghisellini2017}, the blazar sequence posits an inverse relationship between gamma-ray luminosity and peak frequency of the low-energy synchrotron hump for gamma-ray blazars.  Extending the predictions of the blazar sequence to even dimmer blazars suggests that the dimmest gamma-ray blazars would have even higher synchrotron peaks $\nu_{syn} > 10^{15} \:\rm{Hz}$. Though works like \cite{Fossati1998} and \cite{Ghisellini2017} make a physical case for the blazar sequence, the relationship between luminosity and synchrotron peak frequency has been observed for several blazar samples, regardless of any preferred physical interpretation. To this end, the continued study of gamma-ray blazars in the \textit{Fermi} unassociated sources can test the blazar sequence by characterizing dimmer gamma-ray blazars and comparing with the bright blazars used to construct the original blazar sequences.

The blazar sequence is not entirely without controversy. As with many flux-limited samples throughout the history of astronomical research, there is a real possibility of observational biases against certain types of blazars. Some recent arguments suggest that the relationships of the blazar sequence are selection effects \citep{Giommi2012,Giommi2015} leading to samples biased against dim, red blazars. For example, a dim blazar with low peak frequencies might be too dim in gamma-rays to be detected or associated in \textit{Fermi} catalogs, preventing the construction of a sample of red dim blazars. Blazars in the unassociated \textit{Fermi} sources can serve as a valuable sample to test these competing hypotheses; as the unassociated sources have generally lower gamma-ray fluxes than the associated sources, the blazars of the unassociated sources could include some red dim blazars. If the dimmest blazars have a disproportionate mix of high and low $\nu_{syn}$, the general predictions of the blazar sequence may need to be revised.

2157 of the 4FGL sources are labeled as unassociated, lacking confident association with lower-energy counterparts. These gamma-ray sources represent a treasure trove of new objects to test prominent theories of high-energy astrophysics. Given that a plurality of associated sources in 4FGL-DR3 are blazars, extrapolation suggests that there are hundreds if not thousands of blazars in the unassociated sources. If identified and investigated, these blazars can significantly supplement current catalogs of gamma-ray blazars \citep{4LACDR3} with lower-flux blazars to combat flux biases in established samples and test the limits of current blazar theory.

Given the multiwavelength nature of blazars, gathering flux data from as wide a range of energies as possible is an important tool for analyzing blazars in a broadband context.  The blazars of the 4LAC catalog have been extensively surveyed and investigated, many having spectroscopic redshifts, but any blazars found in the unassociated source sample have much sparser coverage. As the unassociated sources often lack targeted observations, the counterparts localized to within $\sim 5 \arcsec$ with \textit{Swift} observations in \citep{Kerby2021} make it feasible to sift through crowded optical, IR, and radio catalogs searching for spatially coincident emission. Because these counterparts are only recently characterized, in-depth population analysis of the unassociated sources is still in its nascent stages.

The position of the $\nu F_\nu$ peaks in the low-energy and high-energy bumps $\nu_{syn}$ and $\nu_{Com}$ plus the ratio of the fluxes at both peaks (the Compton dominance ratio) are relevant quantities that help characterize the physical processes of individual blazars. $\nu_{syn}$ constrains the typical electron energy and magnetic field within the emitting region of the blazar jet.  $\nu_{Com}$, when compared to $\nu_{syn}$, expresses a relativistic boosting factor for upscattering of photons in an SSC model but can also constrain alternative emission models including hadronic and external Compton methods. Finally, investigating Compton dominance gives insights into specifics of the emissions processes of the jet. All three can be obtained from the SED alone without physical modeling, and indeed works like \cite{4LACorig} obtain values for each using polynomial or log-normal fits to broadband SEDs. Physical modeling of jet mechanics, while involving several degenerate parameters requiring in-depth observations to constrain, produces estimates both for physical jet parameters and for phenomenological variables like $\nu_{syn}$. 

In this work, we use two recently published tools to estimate these spectral properties and examine our sample of 106 blazars identified from the \textit{Fermi} unassociated sources (the ``new" blazars). \verb|BLaST| \citep{BLAST2022} uses machine learning trained on almost 4000 known blazar SEDs to predict the positions of the synchrotron peak frequency, being especially resilient against misreading processes like dust or disk emission as jet synchrotron emission. We apply \verb|BLaST| to our sample to obtain one estimate for $\nu_{syn}$. We also conduct SSC modeling of our SEDs using the \verb|agnpy| modeling package \citep{AGNPY2022} with a \verb|sherpa| fitting routine to estimate the physical jet parameters in each of the new blazars. While these fits have numerous degeneracies between physical parameters and fairly large uncertainties, producing reasonable fits to the SED allows for numerical location of peak frequencies and Compton dominance while comparing to the physical parameters of brighter blazars.

In section \ref{sec:Samples} we describe the sample of likely blazars identified from the unassociated \textit{Fermi} sources, and we describe the archival data used to build broadband SEDs of each blazar.  In section \ref{sec:Analysis} we derive spectral properties for each blazar SED, and we conduct physical fitting of the emission mechanisms for each. Finally, in section \ref{sec:Results} we compare our results to established catalogs of brighter blazars in the 4LAC catalog, contextualize our results with respect to current debates in blazar theory, and suggest next steps for further analysis. In this work we adopt a $\Lambda$CDM cosmology with $h=0.7$ and $\Omega_M = 0.3$.

\section{Samples and Spectra} 
\label{sec:Samples}

\subsection{The New Blazar Sample}
\label{sec:Unassoc}

\cite{Kerby2021} (henceforth K21) describes recent results of an ongoing campaign of \textit{Swift}-XRT and -UVOT observations at \textit{Fermi} unassociated sources, producing a list of over 200 such unassociated targets with X-ray and UV/optical counterparts.  Neural network classification trained on established lists of gamma-ray blazars and pulsars showed that 135 of the examined sources were likely blazars, creating a sample of dim blazars with gamma-ray, X-ray, and UV/optical detections. The localization of \textit{Fermi} unassociated sources to within $5 \arcsec$ using X-ray detections, down from the $\sim 2 \arcmin$ uncertainty ellipses in the \textit{Fermi} catalog, is similarly a vital step in further studies of the unassociated sources, allowing for detailed and precise follow-up.

Subsequently, \cite{Kaur2022} (henceforth U22) combined \textit{Fermi} gamma-ray features, \textit{Swift} spectral parameters, and WISE colors to classify the new blazar sample into likely BL Lac or FSRQ categories. The neural network classification in U22 suggested that the majority of our sample of new blazars were more similar to BL Lacs than FSRQs. Collected results and cross-references in U22 showed that the new blazars have properties and redshifts at first glance moderately similar to known blazars, a starting point for this more in-depth analysis of the new blazars. The 106 new blazars in U22 is the foundation of the sample used in this work, reduced from the 135 likely blazars detailed in K21 by requiring WISE IR counterparts.

\begin{figure}[h]
    \centering
    \includegraphics[width=\columnwidth]{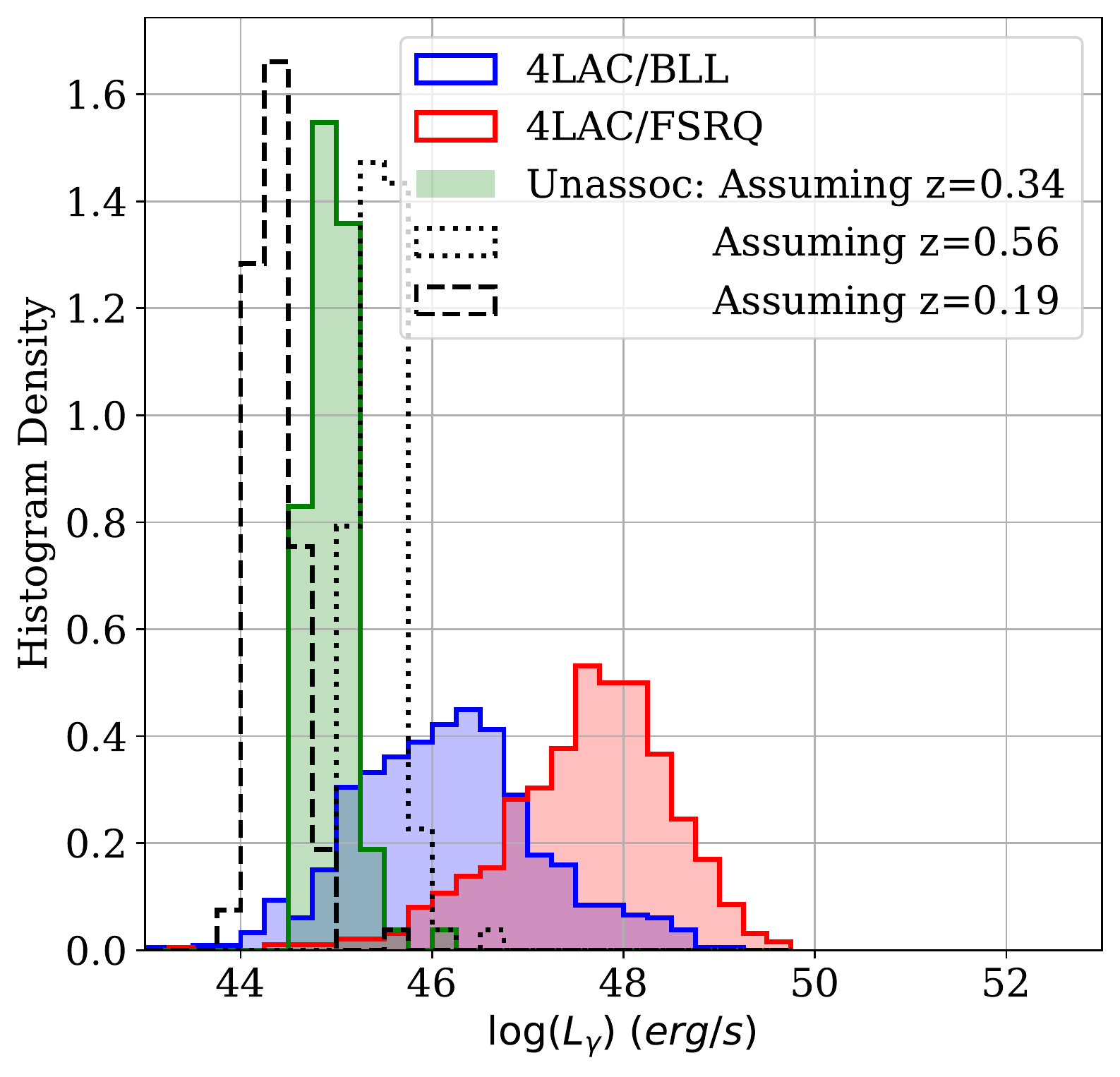}
    \caption{Comparing the gamma-ray luminosities of 4LAC BL Lacs (blue) and FSRQs (red) with estimated luminosities for the new blazars (green) assuming the same median redshift as the BL Lac sample $z=0.34$.}
    \label{fig:LuminHist}
\end{figure}

Using only the \textit{Fermi} gamma-ray data for the unassociated sources, it is possible to do a rudimentary comparison with the known blazars of the 4LAC catalog by making a few assumptions. Applying the median redshift of the BL Lacs in the 4LAC catalog \citep{4LACDR3} $z=0.34$ to the new blazars assumes that the new blazars have the same redshift distribution as the brighter 4LAC BL Lacs. In this way, we can estimate the gamma-ray luminosities of the new blazars if they are within the same redshift range as the 4LAC BL Lacs. The histogram in Figure \ref{fig:LuminHist} shows the gamma-ray luminosity of the 106 new blazars (green) assuming $z=0.34$ compared to the 4LAC BL Lacs (blue) and FSRQs (red) with known redshifts. Since our sample of new blazars certainly has a diverse range of redshifts, we also plot two additional histograms representing the luminosity estimates if the new blazars are at the 25th and 75th percentiles of the BL Lac redshift distribution to illustrate a range of possible luminosities.

Comparing the luminosity distributions, it is clear that if the new blazars have roughly the same redshift distribution as the known BL Lacs, then they are intrinsically much lower luminosity than the 4LAC BL Lacs, and of course even dimmer still than the 4LAC FSRQs. Alternatively, if the new blazars are intrinsically the same luminosity as the known BL Lacs, they must be higher redshift. Some combination of these two effects is also feasible, with the new blazars being both higher redshift and lower luminosity. Given the results of K21 and U22, the blazars found in the unassociated source sample are dimmer than the typical BL Lac or FSRQ in the 4LAC catalog and also have properties more like BL Lacs than FSRQs. More detailed spectral analysis tests these and other findings about dim blazars.

\subsection{SED Construction}
\label{sec:SEDCons}

The \textit{Fermi}-LAT point source 12-year catalog \citep[4FGL-DR3,][]{4FGLDR3} includes the gamma-ray fluxes for each unassociated source in eight energy bands, plus upper and lower uncertainties for each band.  Given the lower average flux of the unassociated sources, many of the fluxes in individual energy bands are marginal, with lower error bars extending to zero.  The units of the \textit{Fermi} catalog $\nu F_\nu$ fluxes are $\rm{erg/s/cm^2}$, the required unit for the subsequent analysis techniques, so no conversions are necessary to begin building multiwavelength SEDs for each new blazar. For each energy band, we use the logarithmic center of the range of each energy band converted to $\rm{Hz}$ as the central point.

Building off the \textit{Fermi} fluxes, we incorporate the \textit{Swift}-XRT data used in K21 and U22 into each spectrum.  While K21, U22, and other previous works conducted independent X-ray fitting using power-law models, for this work we use raw photon lists to facilitate physical modeling. Similarly to the \textit{Fermi}-LAT analysis, we sum up the $\nu F_\nu$ flux in each of five logarithmically spaced bins between $1$ and $10 \:\rm{keV}$, using Poisson statistics to infer the uncertainties on the photon flux in each bin. Assuming the galactic $n_H$ values for each source position \citep{Wilms2000}, we rescale the XRT observed flux values to unabsorbed flux.

Though the \textit{Swift}-XRT detector can observe photons with energies as low as $0.3 \:\rm{keV}$, we exclude the lower energy range from $0.3$ to $1 \:\rm{keV}$ due to the dramatic upper and lower error bars produced by a combination of low photon statistics, high flux uncertainties, and significant $n_H$ corrections at low energies. The large error bars proved troublesome for comparing the analysis methods detailed below, and few photons were lost by excluding this lower energy range.

Notably, because the observations used to construct SEDs for these dim blazars draw on a wide range of archival data and are not simultaneous, different energy bands could include observations with the blazar in flaring or quiescent states. Unfortunately, there is little after-the-fact recourse for the problem of non-simultaneous observations for these extremely dim blazars because the summation of over a decade of \textit{Fermi} observations was necessary to obtain gamma-ray detection and launch our multiwavelength analysis.

\begin{figure}[h]
    \centering
    \includegraphics[width=\columnwidth]{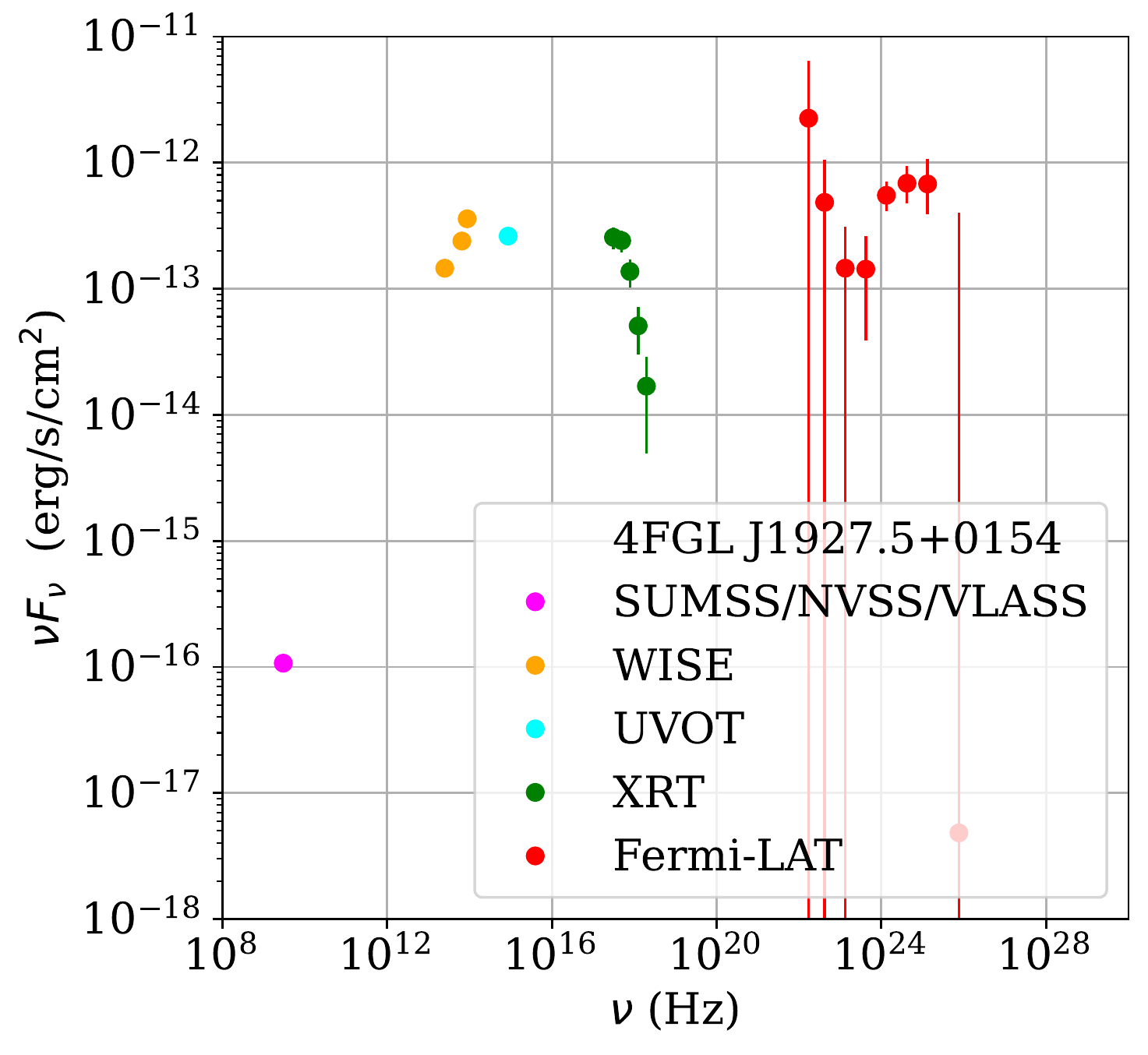}
    \caption{A typical SED for a blazar in our sample. SEDs of all of the 106 new blazars are available in an associated repository, and a small selection of characteristic SEDs is presented at the end of this paper.}
    \label{fig:ExampleSED}
\end{figure}

\textbf{Several of the \textit{Fermi}-LAT and \textit{Swift}-XRT energy bands have lower limits that include zero, as shown in Figure \ref{fig:ExampleSED}. The \textit{Fermi} fluxes and errors are extracted directly from the 4FGL database, precluding rebinning. While the dimmest \textit{Swift}-XRT counterparts have lower limits that reach zero within some specific energy bins, the energy-integrated detection procedure explained in \cite{Kerby2021} ensures that only $\rm{S/N}>4$ XRT sources are included in this work. These lower limits do not interfere with the analysis detailed below.}

All the likely blazars in our sample have UV or optical detections in one or more of the six filters of the \textit{Swift}-UVOT detector. These lower-energy counterparts are described in K21, the magnitudes already corrected for absorption.  For each counterpart, the UVOT AB magnitudes are each converted to $\nu F_\nu$ flux via the standardized relation given in Equation \ref{eq:ABmag}, using the centroid frequency for each UVOT band as the $\nu$ value for placing each magnitude on the SED.

\begin{equation} \label{eq:ABmag}
    \nu F_\nu = \nu \times 10^{\frac{m + 48.6}{-2.5}} \:\rm{erg/s/cm^2}
\end{equation}

In U22, each likely blazar had WISE magnitudes extracted via source cross-match with the WISE all-sky source catalog \citep{Cutri2012} within the positional uncertainty of the \textit{Swift}-XRT detection, $5 \arcsec$. While U22 used WISE colors for neural network classification of likely blazars, we convert the w1, w2, and w3 magnitudes of each likely blazar to AB magnitudes using the magnitude offsets given in Table 8 of Section 4.4h on the WISE data processing website\footnote{\url{https://wise2.ipac.caltech.edu/docs/release/allsky/expsup/sec4_4h.html}}. The AB magnitudes are similarly converted to fluxes via Equation \ref{eq:ABmag}. 

WISE magnitudes are vital for constraining the low-energy side of the synchrotron peak of blazars and for demarcating the synchrotron peak frequency, as in blazars they often occupy the beginning of the power-law slope decreasing to lower energies. Especially for dim blazars, the total emission from AGN-hosting galaxies includes other components that can contribute significantly in the IR band, such as galactic stellar emission.
 
Finally, we conduct a position cross-match between the XRT centroids of the likely blazars with the SUMSS \citep{SUMSS}, NVSS \citep{NVSS}, and VLASS \citep{VLASS} radio surveys. The cross-match again has a tolerance of $5 \arcsec$, the localization of the \textit{Swift}-XRT detections. While these three surveys together cover the entire sky, many of the likely blazars do not have counterparts in the point source lists of these surveys. Given the steep radio slope of many blazars, it is not unexpected that dim blazars like those in our sample can be undetected in the above catalogs. Fluxes in $\rm{mJy}$ for radio counterparts are converted to $\nu F_\nu$ in $\rm{erg/s/cm^2}$.

Combining the gamma-ray through radio fluxes, we produce a SED for each source such as that shown in Figure \ref{fig:ExampleSED}.  After examining each of the 106 likely blazars manually, all show a feasible "two hump" spectrum characteristic of a blazar, typically with radio through X-ray forming the low-energy synchrotron peak and with gamma-ray emission constraining the high-energy peak. The fact that most of the sources have negative spectral indices in the X-ray suggests that the X-ray emission is not located in the trough of the two-hump spectrum or in the high-energy peak, but instead in the synchrotron peak. This is an early indication that many of the likely blazars in our samples have high synchrotron peak frequencies, as the X-ray spectra of FSRQs with low $\nu_{syn}$ often shows positive slope on the low-energy side of the high-energy bump \citep{Ghisellini2017}.

\section{SED Analysis}
\label{sec:Analysis}

\subsection{BLAST Synchrotron Peak Estimation}

The recently released \verb|BLaST| tool \citep{BLAST2022} is a machine learning estimator for predicting the $\nu F_\nu$ synchrotron peak frequency $\nu_{syn}$ given a multiwavelength SED for a blazar. \verb|BLaST| has not yet been independently applied to new samples of blazars, so our multi-pronged approach to SED analysis allows for methodological comparisons on a unique sample. \verb|BLaST|, trained on a sample of 3793 blazars with $\nu_{syn}$ ranging from $10^{12}$ to $10^{18} \:\rm{Hz}$, showed improvements compared to the polynomial fitting of the 4LAC-DR2 \citep{4LACDR2} values for $\nu_{syn}$, especially in blazars with significant IR or optical excesses from non-synchrotron components like disk or dust emission. 

\begin{figure*}[t]
    \centering
    \includegraphics[width=\textwidth]{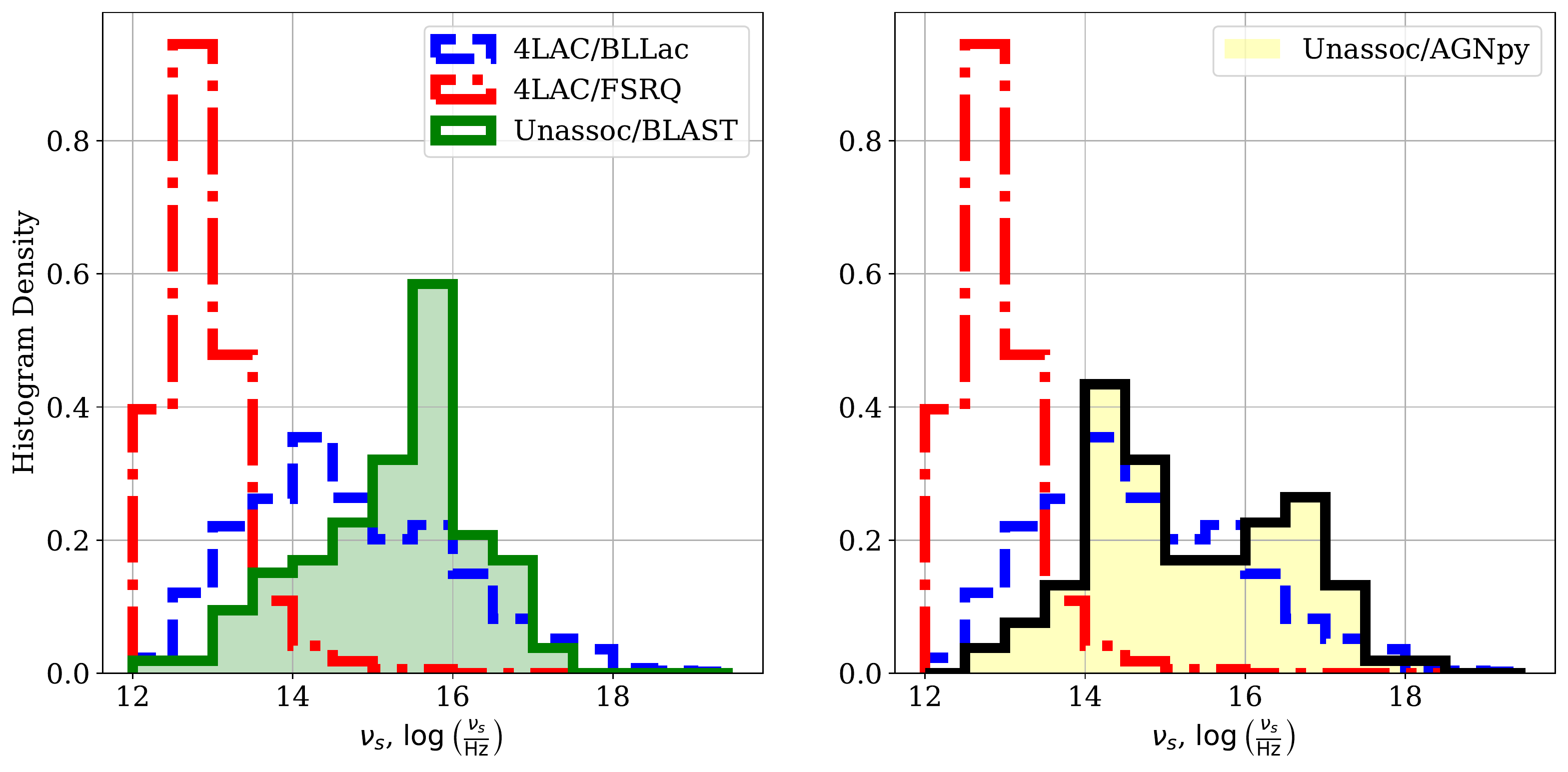}
    \caption{Synchrotron $\nu F_\nu$  peak frequency $\nu_{syn}$ for the new blazars via BLaST (green, left) and agnpy (yellow, right) compared to 4LAC FSRQs (red dash-dot) and BL Lacs (blue dashed). The green and yellow histograms do not show a significant population of new blazars with low $\nu_{syn}$, in agreement with the predictions of the blazar sequence.}
    \label{fig:NuSynCompare}
\end{figure*}

Using the SEDs constructed in Section \ref{sec:SEDCons}, we use \verb|BLaST| to estimate the synchrotron peak frequency for our sample, along with the uncertainty in $\nu_{syn}$.  Given the predictions of the blazar sequence, estimating $\nu_{syn}$ can show if the dim blazars have high synchrotron peaks to extend the blazar sequence or if they have low synchrotron peaks to break from it. The results of the \verb|BLaST| estimations are shown in the left panel of Figure \ref{fig:NuSynCompare}.

With a median $\log (\nu_{syn}/\rm{Hz}) = 15.5$, the new blazars have significantly higher synchrotron peak frequencies than the populations of BL Lacs or FSRQs in the 4LAC catalog. The BL Lacs in the 4LAC catalog have a median $\log (\nu_{syn}/\rm{Hz}) = 14.2$ while the 4LAC FSRQs have a median $\log (\nu_{syn}/\rm{Hz}) = 12.7$, suggesting that few if any of our new blazars are similar to FSRQs and that they tend to have rather high $\nu_{syn}$ even for BL Lacs. Figure \ref{fig:NuSynCompare} shows how distribution of $\nu_{syn}$ for the new blazars mostly overlaps with the high synchrotron-peaked BL Lacs of the 4LAC catalog.

With the \verb|BLaST| estimations for $\nu_{syn}$, it is possible to place the new blazars in the context of the blazar sequence by plotting $\nu_{syn}$ versus $L_{\gamma}$, the gamma-ray luminosity estimated from the median redshift of the BL Lac and FSRQ 4LAC samples. In this way, we replicate Figure 2 from \cite{Finke2013}, a previous result that shows inverse relationship between synchrotron peak frequency and luminosity for BL Lac objects. Conducting this comparison also tests the prediction of U22 that the new blazars are more like BL Lac objects than FSRQs.

\begin{figure}[h]
    \centering
    \includegraphics[width=\columnwidth]{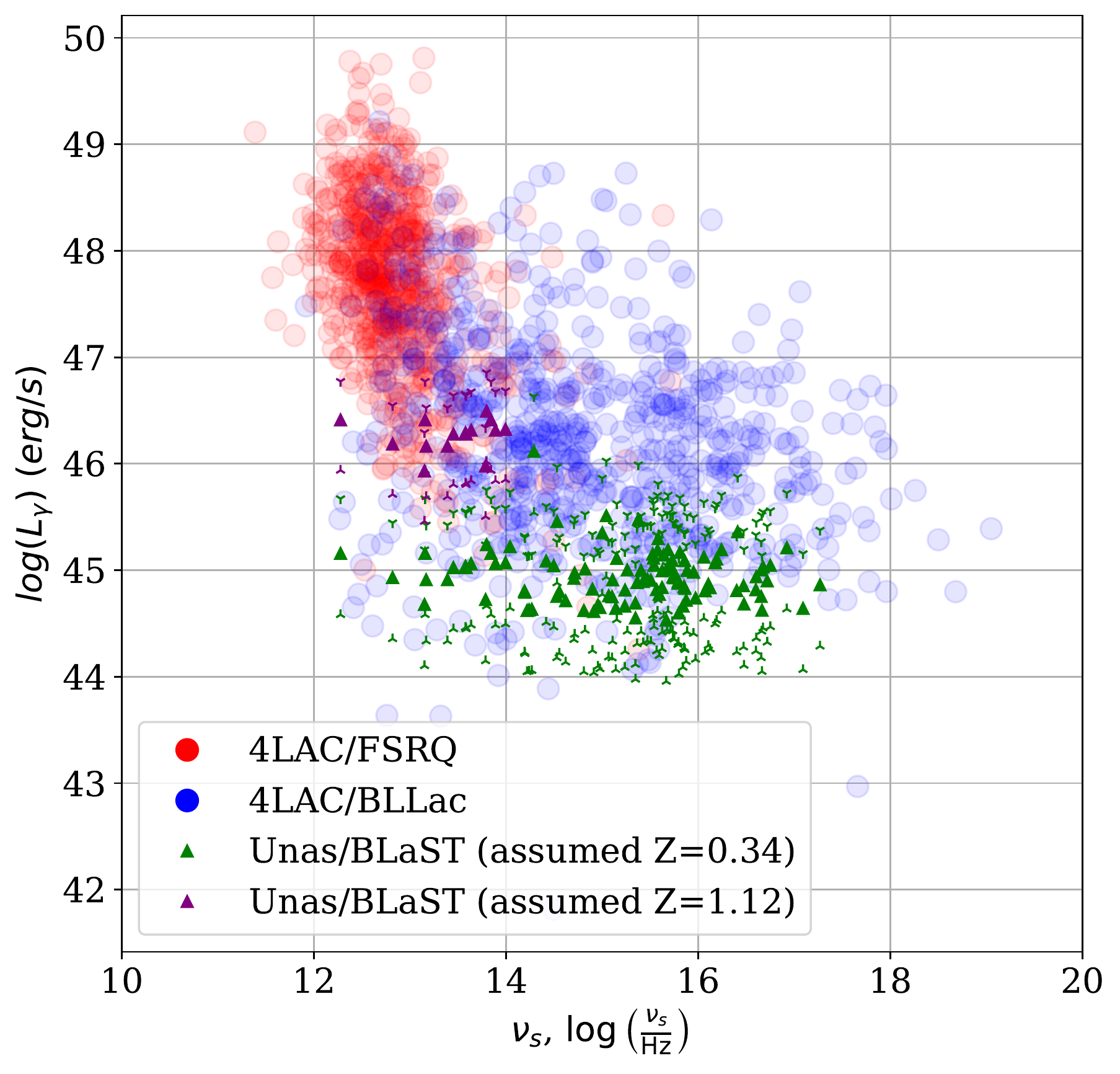}
    \caption{Synchrotron $\nu F_\nu$  peak frequency $\nu_{syn}$ via BLAST versus gamma-ray luminosity for the 4LAC FSRQs (red, transparent), BL Lacs (blue, transparent), and new blazars. Green points and arrows represent the new blazar luminosities using 25th, 50th, and 75th percentiles of the 4LAC BL Lac redshift distribution. Purple points represent the same, using the percentiles for the 4LAC FSRQs}
    \label{fig:BlastBlazSeq}
\end{figure}

Figure \ref{fig:BlastBlazSeq} superimposes the new blazars onto the brighter BL Lacs and FSRQs from the 4LAC catalog. Applying the 25th, 50th, and 75th percentiles of redshift from the 4LAC BL Lac sample, we predict the gamma-ray luminosities of the new blazars. For the small subset of our sample of new blazars with low synchrotron peaks, we also predict their luminosities with the percentiles from the 4LAC FSRQ sample, as these blazars may be FSRQs given their $\nu_{syn}$ values.

Clearly, if the new blazars have the same redshift distribution as the known BL LACs, they represent a sample of intrinsically lower luminosity blazars, with median absolute luminosity lower by approximately an order of magnitude. Likewise, placing the new FSRQ blazars at the same distance as the typical FSRQ population, leads to the result that the new FSRQs are several orders of magnitude lower luminosity than the median 4LAC FSRQ. The lower flux observed from the new blazars could be due to physical jet parameters like a decreased Doppler factor $\delta_D$, which in turn is determined by bulk Lorentz factor or viewing angle. The \verb|BLaST| estimations for $\nu_{syn}$ and its uncertainty are presented in two columns of Table \ref{tab:phenom}.

\subsection{agnpy SSC Modeling}

We also apply the \verb|agnpy| fitting package \citep{AGNPY2022} to our collection of SEDs with three aims. While \verb|BLaST| produces an estimate for $\nu_{syn}$, \verb|agnpy| models physical parameters of the emitting jet. Secondly, \verb|agnpy| allow us to extract phenomenological parameters like Compton dominance ratio, $\nu_{syn}$, and high-energy peak wavelength $\nu_{Com}$ from a modeled spectrum regardless of exact physical parameters. These values allow for direct comparison with established blazars from the 4LAC catalogs, which includes many of these phenomenological parameters. Finally, we can compare the $\nu_{syn}$ predictions of BLaST and \verb|agnpy|, as both tools are relatively new and comparisons can reveal biases or problems yet undetected in both.

To fit the blazar SEDs using \verb|agnpy|, we use a one-zone SSC model for blazar emission, with the electrons in the jet having a broken power law $\gamma$ distribution. While the combination of these models is elementary and neglects various alternative or supplementary emission mechanisms such as hadronic processes or external synchrotron upscattering, it is appropriate for first-look fitting of these dim blazars with few observations. The SSC model implemented in \verb|agnpy| is fully described in \cite{AGNPY2022} and \cite{Finke2008}.

For both \verb|BLaST| and \verb|agnpy|,we included flux errors on each $\nu F_\nu$ data point. \verb|agnpy| uses these one-sigma errors while fitting with the \verb|sherpa| and \verb|gammapy| subroutines, but \verb|BLaST| seems to be unaffected by the flux estimates on individual data points. Indeed, we created a test SED and obtained numerous $\nu_{syn}$ estimates via \verb|BLaST|, with unchanging results regardless of altered errors.  Regardless, lower limits that include zero due to the dimness of our sample (for example, several of the \textit{Fermi}-LAT points in Figure \ref{fig:ExampleSED}) are not impactful for $\nu_{syn}$ estimation either by being seemingly ignored (for \verb|BLaST|) or as less constraining than flux bands with tighter errors (for \verb|agnpy|).

We apply Nelder-Mead \citep{NelderMead1965} optimization with starting parameter estimates and allowable ranges shown in Table \ref{tab:initial}. The broken power law electron distribution has six parameters: break Lorentz factor $\gamma_b$, minimum and maximum electron factors $\gamma_{min}$ and $\gamma_{max}$, power law indices below and above the break $p_1$ and $p_2$, and scaling factor $k_e$.  The SSC model has three further parameters: magnetic field strength $B$, a Doppler factor $\delta_D = \frac{1}{\Gamma (1-\beta \cos{\theta})}$ which depends on bulk Lorentz factor and viewing angle, and variability timescale $t_{var}$. Finally, the entire spectrum has a redshift $z$ which we fix at $z=0.34$, the median redshift of the 4LAC BL Lac blazars. While our sample of dim blazars certainly has a wide range of redshifts, and may have a higher median redshift to partially account for their lower flux, holding the redshift constant during fitting reduces degeneracies between parameters like $t_{var}$ or $\delta_D$ and luminosity distance.  If one should assume that a blazar in our sample has a higher redshift than the median 4LAC BL Lac, other fitted parameters like $\delta_D$ can increase monotonically to maintain the same observed flux.

For the SSC fitting, there are various degeneracies between physical jet parameters in the model, especially when fitting dim blazars that do not have substantial observations constraining redshift and magnetic field.  For example, Equation 21 of \cite{Finke2008} gives the expression for synchrotron emission as

\begin{equation} \label{eq:Sync}
    f_{\epsilon}^{syn} = \frac{\sqrt{3} \delta_D^4 \epsilon^\prime e^3 B}{4 \pi h d_L^2} \int_{-\infty}^{+\infty} d\gamma^\prime N_e^\prime (\gamma^\prime) R(x)
\end{equation}

with $\epsilon^\prime$ the energy of emitted photons in the jet frame, $N^\prime(\gamma^\prime)$ the electron distribution in the jet frame, and $R(x)$ a dimensionless scaling parameter. SSC modeling uses Equation \ref{eq:Sync} to fit each SED, but the first term of Equation \ref{eq:Sync} shows a clear degeneracy between $\delta_D$ and $d_L$, neglecting certain second-order effects. Indeed, by holding all other variables constant and running $\delta_D$ through a range of values, the redshift $z$ monotonically changes to create almost completely identical fits for a given blazar. By holding $z = 0.34$ for our fitting, we shift uncertainty to other parameters like $\delta_D$.

With more detailed observations and analysis such as radio observations or by obtaining photometric/spectroscopic redshifts, $\delta_D$ or $z$ could be independently constrained for our sample of blazars. Without those more detailed observations, it could be appropriate to compare $\delta_D$ and $z$ to brighter blazars by combining them into a composite parameter. Because of this degeneracy and others, our analysis of the fitting results focuses on jet parameters with little degeneracy or on broadband spectral properties like Compton dominance, while more directed future observations will be necessary to exactly constrain all intrinsic properties of our sample of blazars.

\startlongtable
\begin{deluxetable}{cccc}% C for mathmode
\tablecaption{SSC parameter initialization and ranges for agnpy fitting.}
\tablewidth{\columnwidth}
\label{tab:initial}
\tablehead{
\colhead{Parameter} & \colhead{lower} & \colhead{initial} & \colhead{upper} 
}
\startdata
\hline
$\log (k_e)$ & -12 & -4 & 0 \\
$p_1$ & -2 & 2.5 & 4 \\
$p_2$ & 1 & 4 & 8 \\
$\log (\gamma_b)$ & 2 & 5 & 6 \\
$\log (\gamma_{min})$ & 1 & 2 & 4 \\
$\log (\gamma_{max})$ & 5 & 6 & 8 \\
\hline
$z$ \textbf{(fixed)} & & $0.34$ & \\
$\delta_D$ & 1 & 20 & 100 \\
$\log (B/\rm{G})$ & -2 & -1 & 1 \\
$\log (t_{var}/\rm{s})$ & 4 & 5 & 8 \\
\enddata
\end{deluxetable}

Conducting the SSC fits for each blazar in our sample, we present the physical parameters of each fit in Table \ref{tab:agnpyfits}.  Additionally, the SSC fit for each blazar also allows for an independent estimation for $\nu_{syn}$, in yellow on the right panel of Figure \ref{fig:NuSynCompare}. The peak frequencies and fluxes are obtained by finding all the extrema of the \verb|agnpy| SSC fit and then evaluating which maxima are the peaks of the synchrotron and high-energy peaks. Using these located peaks, we estimate the Compton dominance ratio using

\begin{equation}
    CD = \frac{L_{peak}^{Com}}{L_{peak}^{syn}} = \frac{F_{peak}^{Com}}{F_{peak}^{syn}}
\end{equation}

\noindent and the characteristic Doppler boost

\begin{equation}
    \gamma_{boost} = \left(\frac{3 \nu_{Com}}{4 \nu_{syn}} \right)^{1/2}
\end{equation}

These quantities are features of the shape of the spectral fit; though highly dependent on the physical processes of the jet, they can be estimated through simple polynomial or log-normal fitting of the spectrum.  The 4LAC catalog \citep{4LACDR3} estimates these parameters for hundreds of known BL Lacs and FSRQs using non-physical methods, but our analysis of \verb|agnpy| fits allows for direct comparison between our sample of new blazars and established catalogs of known blazars while examining underlying physical parameters. The phenomenological parameters of the \verb|agnpy| spectral fits, plus the bolometric flux obtained by integrating under the fits, are presented in Table \ref{tab:phenom} and are available as machine-readable databases.

\section{Results and Conclusions}
\label{sec:Results}

\subsection{BLaST vs. agnpy Comparison}

\begin{figure}[h]
    \centering
    \includegraphics[width=\columnwidth]{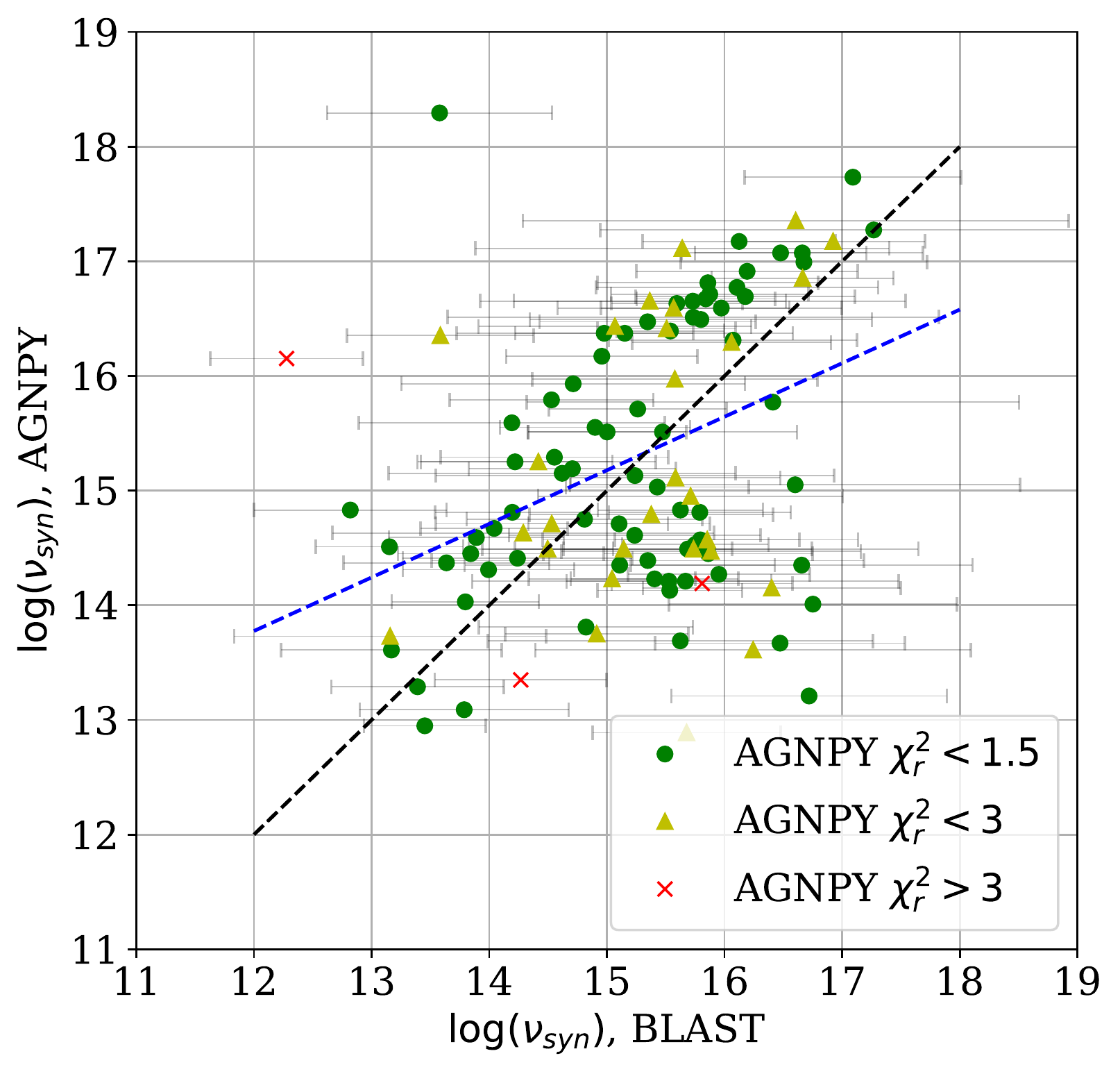}
    \caption{Comparing $\nu_{syn}$ estimates from BLaST and agnpy methods. Points are color-coded based on angpy fit $\chi^2_{r}$. The dashed black line is a one-to-one correspondence, while the dashed blue line is a linear regression to the points with $\chi^2_{r} < 1.5$}
    \label{fig:MethodCompare}
\end{figure}

Because both \verb|BLaST| and \verb|agnpy| produce estimates for $\nu_{syn}$, it is useful to compare the two approaches, especially as both have not yet been widely used.  Figure \ref{fig:MethodCompare} plots the $\nu_{syn}$ estimates from both methods, colored by the goodness-of-fit $\chi^2_{r} = \chi^2 / \rm{D.o.F.}$ in the \verb|agnpy| fitting.  Only \verb|BLaST| returns an uncertainty for $\nu_{syn}$ as it is directly fitting for $\nu_{syn}$ using machine learning, while the value for $\nu_{syn}$ extracted from \verb|agnpy| is a phenomenological estimation form the shape of the entire fitted spectrum.

Though the distribution of $\nu_{syn}$ and the blue regression line show a positive correlation, there is some scatter and the slope of the best-fit line to the two estimates is not unity, the expected result if the two estimates produced identical values for $\nu_{syn}$. Notably, there seem to be a small number of blazars with high $\nu_{syn}$ estimated by \verb|BLaST| but low $\nu_{syn}$ via SSC fitting. Overall, Figure \ref{fig:NuSynCompare} shows that \verb|agnpy| tends to prefer lower $\nu_{syn}$ values for individual blazars compared to \verb|BLaST|.

The most straightforward explanation for this difference is that \verb|agnpy| fits a model including only synchrotron and synchrotron-self-Compton components, while \verb|BLaST| was constructed to account for IR or optical contributions from other sources like dust, the disk, or galactic stars.  If a particular blazar has an IR or optical excess separate from its synchrotron emission, then our simple implementation of \verb|agnpy| would not account for that discrepancy, while \verb|BLaST| would. This would lead to higher $\nu_{syn}$ estimates from \verb|BLaST| via discounting IR or optical contributions from non-synchrotron processes.

Overall, both \verb|BLaST| and \verb|agnpy| SSC fitting are capable tools for obtaining estimates for $\nu_{syn}$. While \verb|agnpy| requires significantly more computation time for multidimensional fits to SEDs, its results can also serve to estimate phenomenological parameters like $\nu_{Com}$ and Compton dominance while constraining physical parameters. However, \verb|BLaST| is trained to avoid deception by IR or optical excesses from non-synchrotron components. The two methods are complementary, especially for investigations focused on obtaining $\nu_{syn}$ values.

\subsection{Comparisons with Bright Blazars}

Both \verb|agnpy| and \verb|BLaST| estimations for $\nu_{syn}$ show that the new blazars tend to have higher synchrotron peaks characteristic of ISP or HSP BL Lac objects. The histograms in Figure \ref{fig:NuSynCompare} show that the \verb|BLaST| (in green) and \verb|agnpy| (yellow) distributions for $\nu_{syn}$ trend higher than the overall BL Lac sample from the 4LAC catalog, and have far greater $\nu_{syn}$ than the FSRQs from 4LAC.  While the physical fitting of blazar SEDs via \verb|agnpy| produced lower $\nu_{syn}$ estimates than \verb|BLaST|, this may be due to the implicit spectral shape invoked in SSC fitting or the lack of accommodation for non-synchrotron contributions to the low-energy peak.

The blazar sequence described in \cite{Fossati1998} and \cite{Ghisellini2017} predicts an anticorrelation between gamma-ray luminosity and $\nu_{syn}$, and our \verb|BLaST| and \verb|agnpy| results support this prediction for a population of dim blazars that have until now been entirely excluded from the construction of the blazar sequence.

\begin{figure}[h]
    \centering
    \includegraphics[width=\columnwidth]{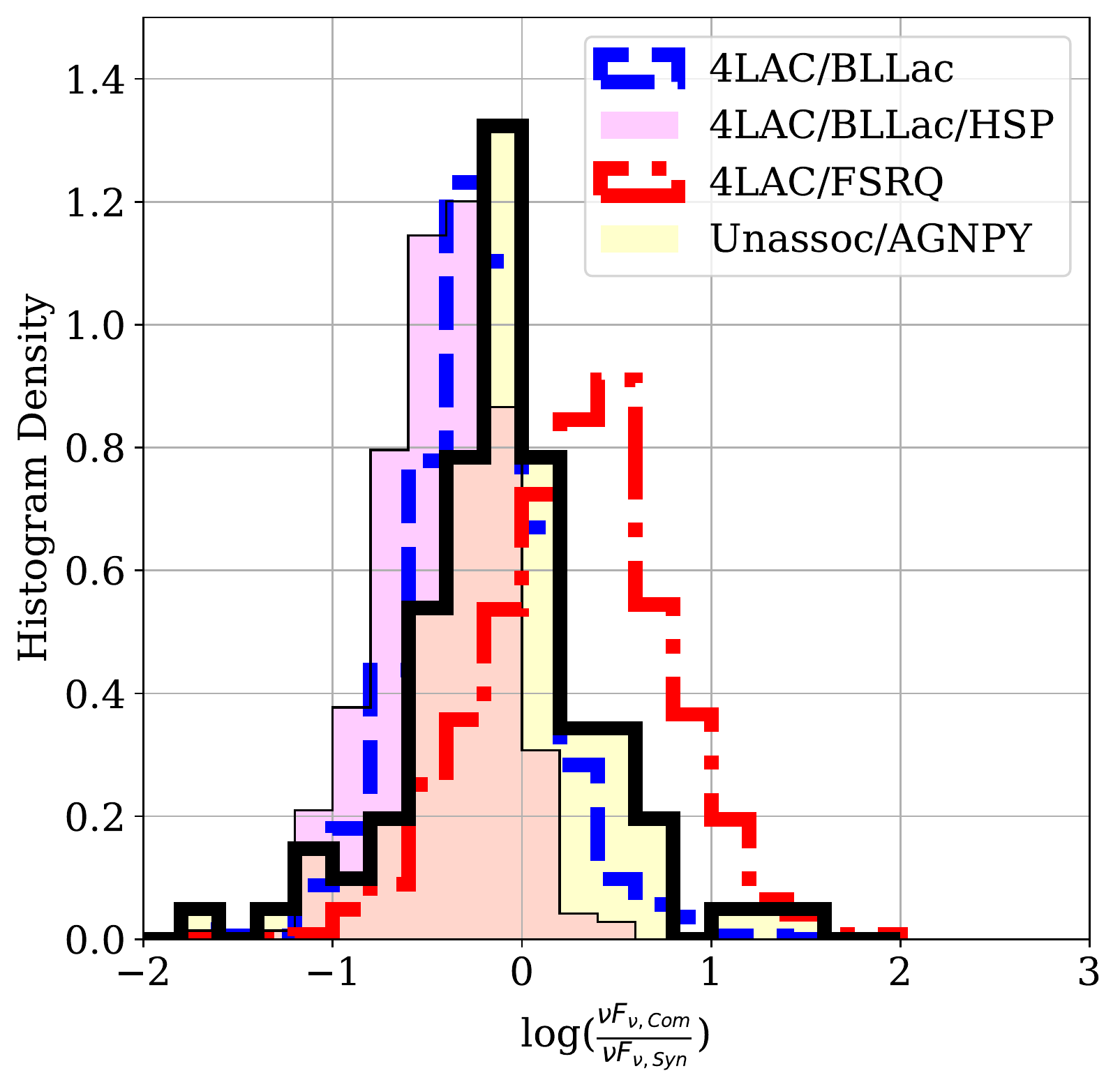}
    \caption{Compton dominance ratio for the new blazars (green) and the 4LAC FSRQs (red) and BL Lacs (blue), plus for 4LAC HSP BL Lacs (faint pink)}
    \label{fig:ComDom}
\end{figure}

The Compton dominance of a blazar is another key parameter of the blazar sequence, evolving alongside luminosity and $\nu_{syn}$ in that framework. \cite{Finke2013} showed that absent absolute luminosity data, a version of the blazar sequence can be constructed by comparing Compton dominance ratio and synchrotron peak frequency, an approach well-suited for our sample of new, dimmer blazars lacking photometric or spectroscopic redshifts. After using the \verb|agnpy| fits to estimate the Compton dominance ratio for the new blazars, we compare with the known BL Lacs and FSRQs of the 4LAC catalog, as well as with a more restricted HSP ($\nu_{syn} > 10^{15} \:\rm{Hz}$) BL Lac subsample. Figure \ref{fig:ComDom} compares the distributions of these blazar samples.

The histograms in Figure \ref{fig:ComDom} show that the Compton dominance ratios of the new blazars are slightly higher than for the \textit{Fermi} BL Lacs in general and the HSP BL Lacs specifically. Conducting a two-sample T-test for equal means, we find the p-value by comparing the Compton dominance ratios of the 4LAC BL Lacs and our sample of new blazars. Obtaining a p-value of 0.007, we find evidence against the null hypothesis that the samples of blazar Compton dominance are drawn from the same population at the 99\% threshold. The slightly higher Compton dominance distribution of our sample of new blazars represents an inversion of the expected trend of brighter FSRQs and BL Lacs.

\cite{Finke2013} showed that the blazar sequence can be reframed in terms of an anticorrelation between $\nu_{syn}$ and Compton dominance, suggesting that our new blazars seem to be a break from this trend, as their Compton dominance distribution is \textit{higher} with their greater $\nu_{syn}$. However, one intriguing prediction of \cite{Finke2013} may shed light on this tension. After physically modelling the SSC emission for blazars and predicting changes in Compton dominance along the blazar sequence, Figure 7 of \cite{Finke2013} shows that for particularly high $\nu_{syn}$ blazars the theoretical Compton dominance ratio should increase as $B$ and the energy density are increased. Because the new blazars studied in this work have notably higher $\nu_{syn}$, the unexpected result in Figure \ref{fig:ComDom} may be an indication of the inverted trend noted in \cite{Finke2013}.

\begin{figure}[h]
    \centering
    \includegraphics[width=\columnwidth]{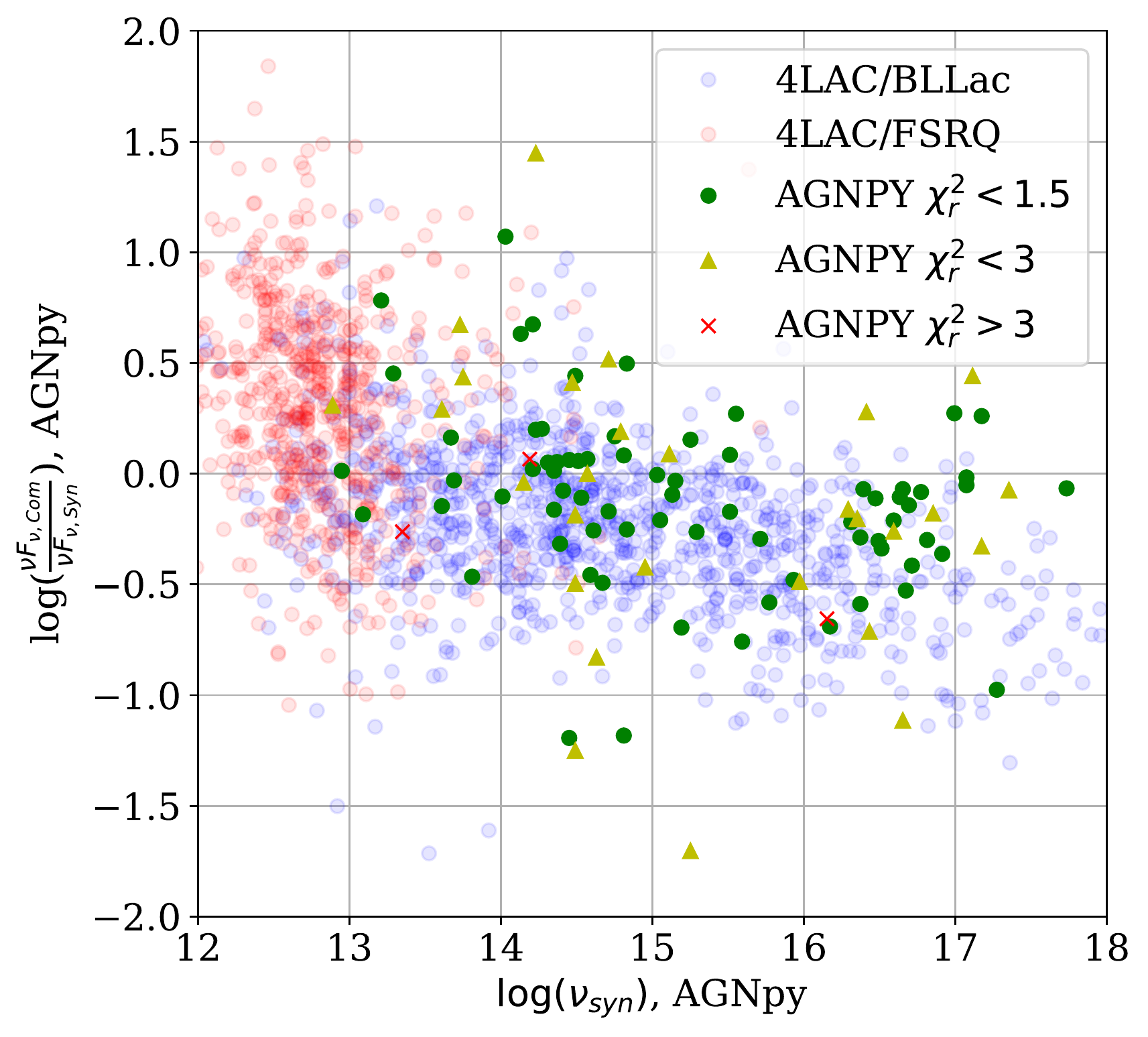}
    \caption{Compton dominance ratio for the new blazars color-coded by SSC fit reduced $\chi^2$, and the 4LAC FSRQs (faint red) and BL Lacs (faint blue). This figure reproduces Figure 5 in \cite{Finke2013}, showing that the new blazars fall within the expected range for BL Lac objects while having lightly higher Compton dominance ratios at high $\nu_{syn}$.}
    \label{fig:FinkeSequence}
\end{figure}

Figure \ref{fig:FinkeSequence} reproduces Figure 5 from \cite{Finke2013}, with the new blazars superimposed on the 4LAC FSRQs and BL Lacs.  While the new blazars tend to fall within the expected phase space in terms of Compton dominance ratio and synchrotron peak frequency estimated by \verb|agnpy|, this comparison demonstrates that the new blazars are similar to ISP and HSP BL Lac objects, occupying the same phase-space of this version of the blazar sequence while having slightly higher Compton dominance ratios (especially at higher $\nu_{syn}$). Further broadband observations of the new blazars could constrain the Compton dominance ratio while testing whether SSC emission models are appropriate for this sample. Even without precisely constrained redshifts, this result shows that our sample of new blazars extends the blazar sequence in terms of Compton dominance and $\nu_{syn}$.

\begin{figure}[h]
    \centering
    \includegraphics[width=\columnwidth]{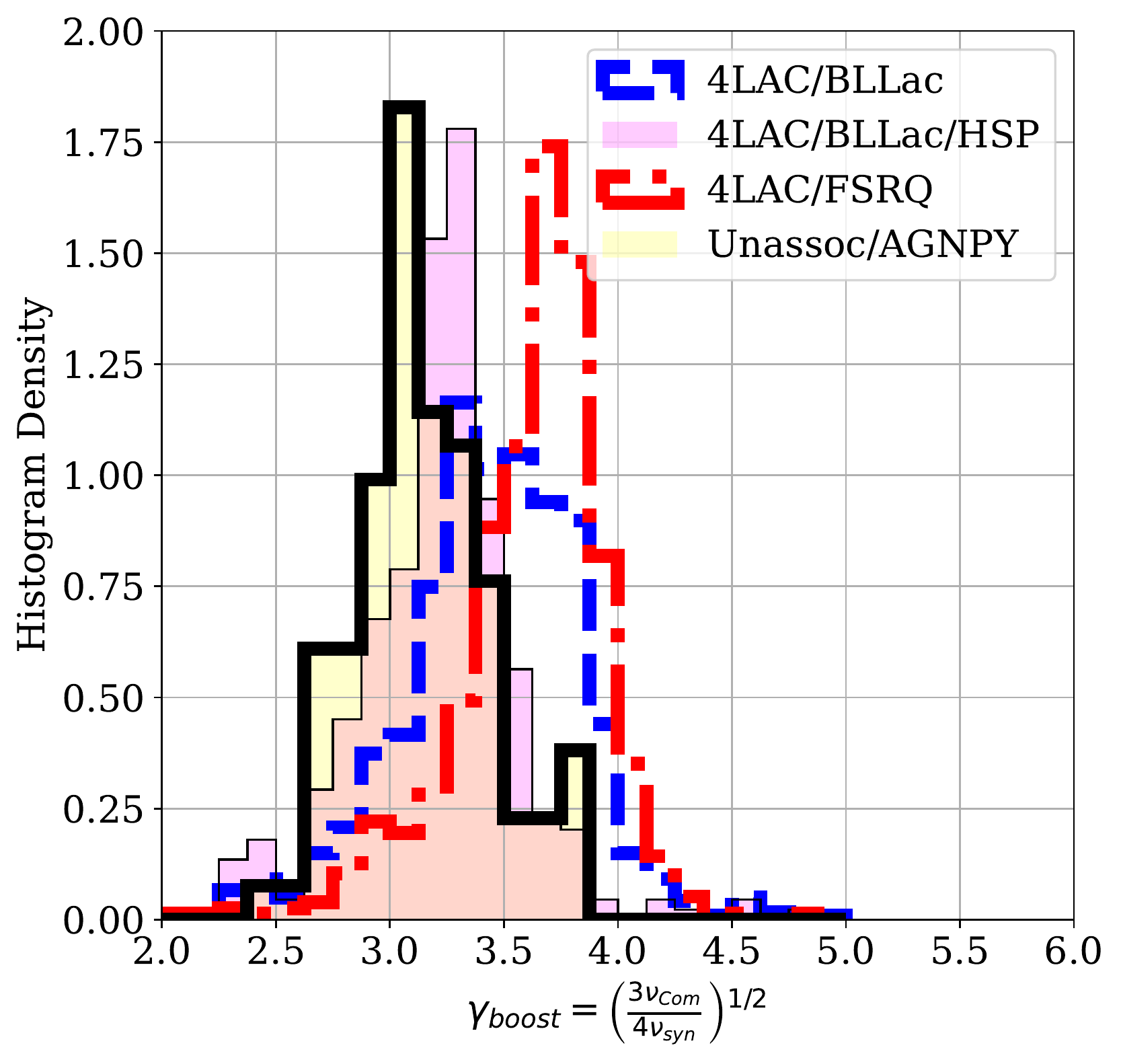}
    \caption{Characteristic Doppler boost $\gamma_{bo}$ for the new blazars (green) and the 4LAC FSRQs (red) and BL Lacs (blue), plus for 4LAC HSP BL Lacs (faint pink). In agreement with the predictions of the blazar sequence, the dim blazars of the unassociated sources have sequentially lower $\gamma_{bo}$ than the HSP BL Lacs of the 4LAC catalog.}
    \label{fig:DoppBoo}
\end{figure}

The characteristic Doppler boost $\gamma_{bo} = \left(\frac{3\nu_{Com}}{4\nu_{syn}}\right)^{1/2}$ between the synchrotron and high-energy peak in a blazar's two-hump spectrum traces a characteristic energy of the emitting electron distribution. While this parameter is a fitted physical parameter of the electron distribution, we measure it directly from the peak frequencies of each blazar spectrum. Figure \ref{fig:DoppBoo} compares the distributions of $\gamma_{bo}$ calculated from the peak positions obtained with \verb|agnpy| to values for the known blazars of the 4LAC catalog. In agreement with the predictions of the blazar sequence framework, this phenomenological approach to measuring $\gamma$ factors in an SSC emission model shows that the new blazars have sequentially lower $\gamma_{bo}$ compared to the HSP BL Lacs, the BL Lacs in general, and the 4LAC FSRQs.

In spectral terms, this means that the synchrotron and high energy peak frequencies of the new blazars tend to be closer than in brighter blazars. This trend suggests that there may be astrophysical differences in the emission regions of our sample of new blazars compared to the brighter 4LAC blazars, and that our examination of the unassociated sources is revealing blazars that extend known blazar samples not only to lower flux but also to new extrema of astrophysical properties.

\begin{figure*}
    \includegraphics[width=\textwidth]{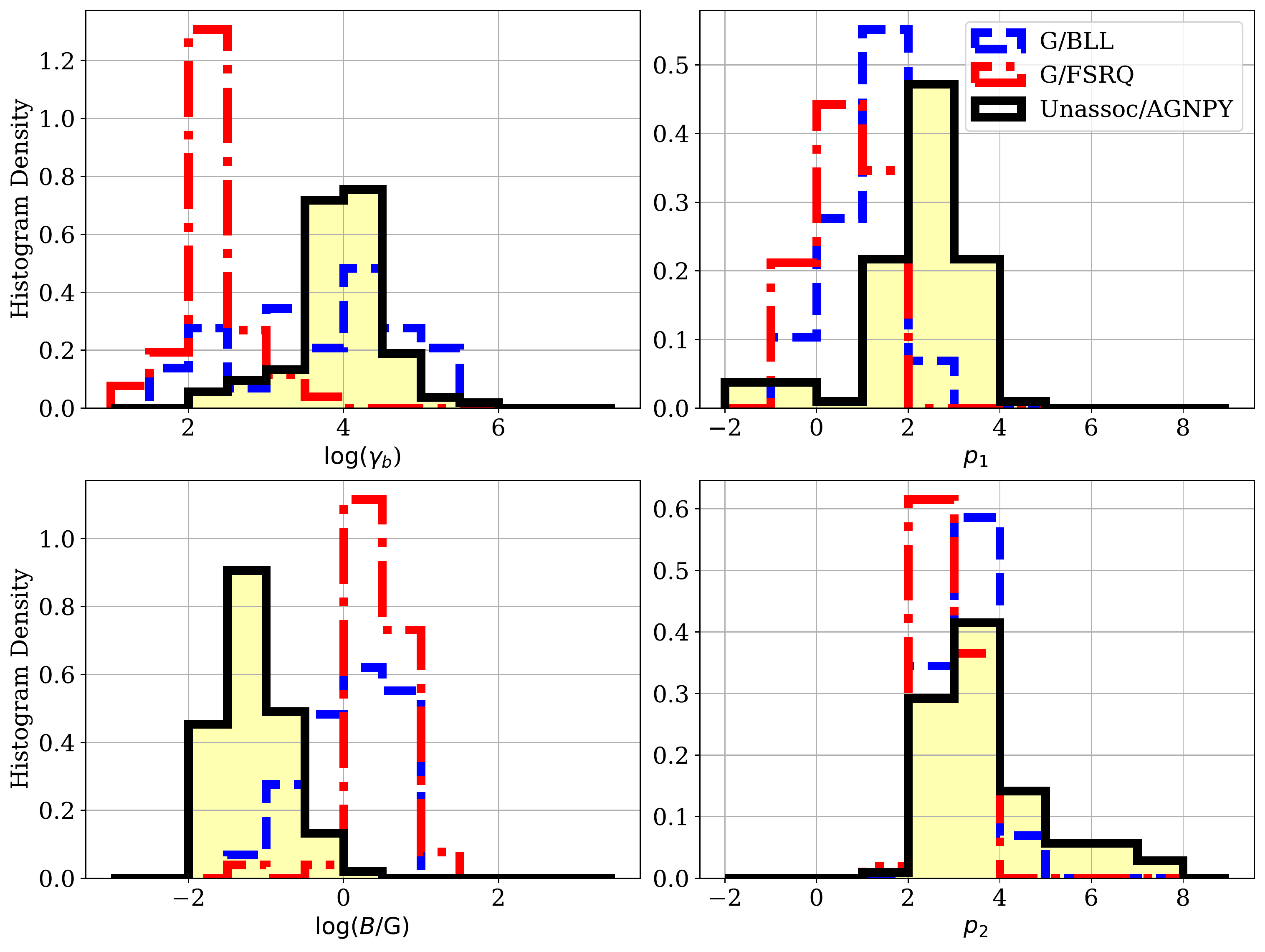}
    \caption{Physical parameters of the emitting electron distribution for the new blazars via agnpy (yellow) and a sample of BL Lacs and FSRQs (blue and red) from \cite{Ghisellini2010}. Clockwise from top left, the logarithm of $\gamma_b$ the break energy in the broken power law electron distribution, $p_1$ and $p_2$ the slopes before and after the break energy, and the logarithm of ambient magnetic field in Gauss.}
    \label{fig:EleDistHist}
\end{figure*}

Using the SSC fits in \cite{Ghisellini2010} for a collection of bright blazars, we can compare physical jet parameters like $B$ and electron distribution parameters with a sample of brighter blazars.  Though the sample in \cite{Ghisellini2010} is small, it contains some of the best-studied bright blazars in the sky modeled with SSC fits similar to our \verb|agnpy| approach, facilitating direct comparison.  In Figure \ref{fig:EleDistHist} we compare fitted parameters of the emitting electron distribution for the new blazars (green) and the BL Lac and FSRQ samples (blue dotted and red dashed). $\gamma_b$ is related to the distance between the two peak frequencies of the two-hump blazar distribution, but here we show the \verb|agnpy| fitted values for completeness.

In terms of $\gamma_b$ and the two slopes of the electron distribution $p_1$ and $p_2$, the new blazars are reasonably similar to the brighter BL Lac sample from \cite{Ghisellini2010}. The magnetic field strength $B$ shows more of a divergence, with the new blazars having generally lower field strength than either the known BL Lacs or FSRQs. This is similarly in agreement with the predictions of the blazar sequence described in \cite{Ghisellini2017}, and together with a lower Doppler factor $\delta_D$ can explain the dimness of the new blazars.  Together with the difference in peak frequency separation in Figure \ref{fig:DoppBoo}, these two plots hint at astrophysical differences in the new blazars beyond continuations of the blazar sequence.
 
\subsection{Summary and Next Steps}

Building SEDs for a sample of 106 blazars identified among the \textit{Fermi} unassociated sources in K21, we used the recently released \verb|BLaST| tool to predict $\nu_{syn}$ for each blazar.  Compared to the brighter blazars of the 4LAC catalog, the new blazars have higher synchrotron peaks shown in Figure \ref{fig:NuSynCompare}, characteristic of HSP BL Lacs. This finding is in agreement with the prediction of the blazar sequence framework \citep{Ghisellini2017} that dim gamma-ray blazars will have high synchrotron peaks. 

It bears noting that the alternative interpretation of the blazar sequence described in \cite{Giommi2012} argues that dim, red blazars are excluded from gamma-ray samples due to selection effects. While the new blazars investigated in this work do tend to have high synchrotron peaks in line with predictions of the blazar sequence, red dim blazars might have low enough gamma-ray fluxes to be undetected (not just unassociated) in the \textit{Fermi} point source catalog. While our findings suggest that the blazars of the unassociated sources tend to have high $\nu_{syn}$, we cannot rule out samples of even dimmer, redder blazars that would be entirely absent from the \textit{Fermi} point source catalog. Still, our results extend the general findings of \cite{Ghisellini2017} to even dimmer \textit{Fermi} blazars, showing that the blazar sequence prediction of anticorrelation between luminosity and $\nu_{syn}$ holds for a sample of blazars previously excluded from the construction of that theory.

With \verb|agnpy| conducting SSC fits of the blazar SEDs with fixed reshift $z = 0.34$ (the median BL Lac redshift in the 4LAC catalog), we obtained estimates for Compton dominance ratio and characteristic relativistic boost $\gamma_{bo}$. Comparing these phenomenological parameters with brighter blazars, we found that the new blazars continue the trends in magnetic field strength $B$ and characteristic Doppler boost $\gamma_{bo}$ predicted in the blazar sequence. However, we also find that the new blazars have unexpectedly high Compton dominance ratios, a result that may be related to the inverted Compton dominance trend predicted in Figure 8 of \cite{Finke2013} for BL Lacs with particularly high $\nu_{syn}$.

Fitting the blazar SEDs with \verb|agnpy| also produces estimates for the physical parameters of the emitting jet and host galaxy for each blazar. While degeneracies between physical parameters make tight constraints on the emitting region difficult, comparing jet parameter estimates to SSC fits of brighter blazars \cite{Ghisellini2010} show that the new blazars have broadly similar physical parameters to brighter blazars, shown in Figure \ref{fig:EleDistHist}. The magnetic field strength $B$ does seem to be slightly lower for the new blazars than for the brighter blazars, another hint at physical differences between the 4LAC blazars and these new blazars from the unassociated sample. Alternatively, the SSC emission model used for \verb|agnpy| fitting may not be the best phenomological model for these dim blazars; while many bright blazars are adequately described by SSC, further observations at these dim blazars may characterize their spectra enough to allow for comparisons with proton cascade \citep{Mannheim1993}, external Compton \citep{Sikora1994}, or proton synchrotron \citep{Mucke2001,Mucke2003} models. An alternative emission mechanism may also explain the noticeably higher Compton dominance ratio shown in Figure \ref{fig:ComDom}.

Further targeted observations could constrain physical parameters of individual blazars and firmly establish their relationship with the wider blazar population. Establishing photometric or spectroscopic redshifts of a subset of our low-luminosity blazar sample would firmly constrain the absolute gamma-ray luminosity of each object and more confidently place each in the context of the blazar sequence. Given that our results suggest that many of these blazars are BL Lac objects, it may be difficult to obtain spectroscopic redshifts of such dim sources lacking optical lines.

We chose an SSC model to fit the blazar SEDs for its simplicity and its phenomenological flexibility, but for many brighter blazars more complex models are necessary. Investigating new blazars in our sample that are not well fit by the SSC model can be used to double-check our sample for incorrect inclusions while allowing for exploring alternative emission mechanisms.  Just as K21 investigated unusual or unique unassociated sources from the wider catalog, the SEDs constructed in Section \ref{sec:SEDCons} are a starting point for individual analysis. Similarly, creating SEDs of unassociated targets that fit neither blazar nor pulsar classification in K21 would clarify the nature of the remaining unclassified sources. 

\software{Astropy \citep{Astropy}, numpy \citep{NumPy}, Matplotlib \citep{Matplotlib}, FTools \citep{FTOOLS}, agnpy \citep{AGNPY2022}, BLaST \citep{BLAST2022}}

\acknowledgments

This research has made use of data and/or software provided by the High Energy Astrophysics Science Archive Research Center (HEASARC), which is a service of the Astrophysics Science Division at NASA/GSFC. We gratefully acknowledge the support of NASA grants 80NSSC17K0752 and 80NSSC18K1730. Finally, we are grateful for the helpful comments and suggestions of our anonymous reviewer.

\bibliography{4FGL}{}

\begin{figure*} \label{fig:Appendix}
\centering
\begin{tabular}{cc}
 \includegraphics[height=7cm]{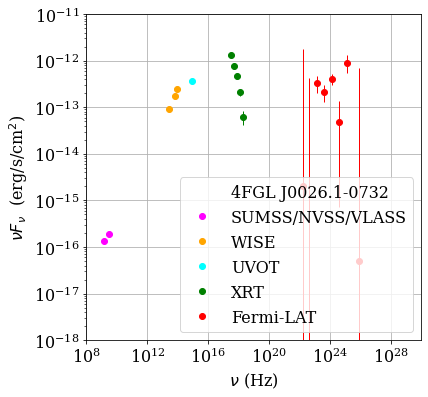} & \includegraphics[height=7cm]{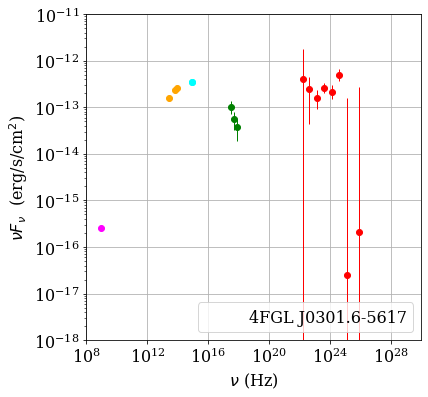} \\
 \includegraphics[height=7cm]{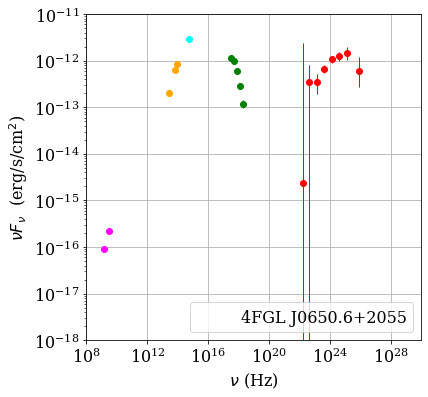} & \includegraphics[height=7cm]{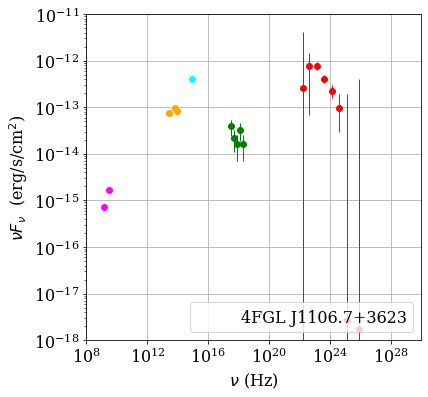} \\
 \includegraphics[height=7cm]{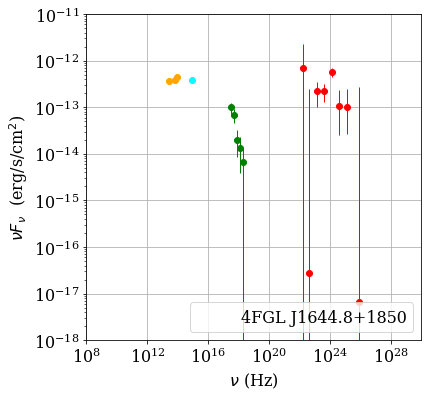} & \includegraphics[height=7cm]{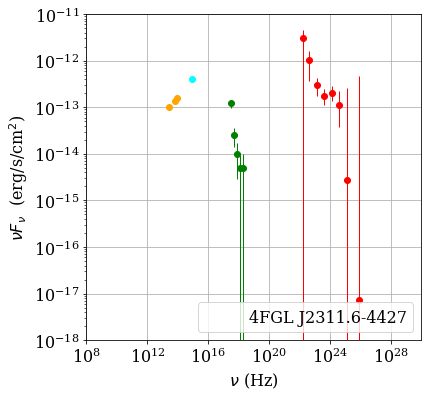} \\
\end{tabular}
\caption{Representative SEDs for a small selection of the 106 blazars examined in this work.}
\end{figure*}

\begin{longrotatetable}
\begin{deluxetable*}{cccccccccc}% C for mathmode
\tablecaption{Phenomological features of the broadband SEDs from agnpy fitting unless otherwise noted \label{tab:phenom}}
\tablewidth{\columnwidth}
\tablehead{
\colhead{4FGL} & \colhead{XRT} & \colhead{$\log(\nu_{syn,BLAST}$)} & \colhead{$\log(\delta \nu_{syn,BLAST})$} & \colhead{$\log(\nu_{syn})$} & \colhead{$\log(\nu F_{\nu,syn})$} & \colhead{$\log(\nu_{Com})$} & \colhead{$\log(\nu F_{\nu,Com})$} & \colhead{$\log(F_{bol})$} \\
\colhead{} & \colhead{} & \colhead{$\log(\rm{Hz})$} & \colhead{$\log(\rm{Hz})$} & \colhead{$\log(\rm{Hz})$} & \colhead{$\log(\rm{erg/s/cm^2})$} & \colhead{$\log(\rm{Hz})$} & \colhead{$\log(\rm{erg/s/cm^2})$} & \colhead{$\log(\rm{erg/s/cm^2})$}
}
\startdata
J0004.4-4001 & J000434.1-400036 & 14.707 & 0.626 & 15.191 & -11.728 & 23.7 & -12.423 & -9.545 \\
J0025.4-4838 & J002536.8-483808 & 15.794 & 1.326 & 14.571 & -12.304 & 22.519 & -12.238 & -9.931 \\
J0026.1-0732 & J002611.9-073115 & 16.6 & 0.829 & 15.051 & -12.444 & 23.84 & -12.654 & -10.01 \\
J0031.5-5648 & J003135.1-564640 & 15.595 & 1.047 & 16.633 & -12.51 & 24.601 & -12.615 & -10.058 \\
J0037.2-2653 & J003729.6-265043 & 13.787 & 1.759 & 13.089 & -12.449 & 21.297 & -12.633 & -9.975 \\
J0057.9+6326 & J005758.1+632642 & 15.678 & 0.708 & 12.889 & -12.545 & 21.197 & -12.237 & -9.693 \\
J0058.3-4603 & J005806.4-460417 & 17.091 & 0.821 & 17.734 & -12.773 & 24.801 & -12.839 & -10.215 \\
J0118.3-6008 & J011823.10-600753 & 14.24 & 1.155 & 14.411 & -12.248 & 23.119 & -12.325 & -9.916 \\
J0120.2-7944 & J011914.7-794509 & 15.861 & 1.318 & 14.451 & -12.586 & 22.679 & -12.524 & -10.171 \\
J0125.9-6303 & J012548.3-630244 & 16.073 & 0.771 & 16.312 & -12.599 & 24.401 & -12.817 & -10.123 \\
J0156.3-2420 & J015624.4-242004 & 15.712 & 0.712 & 14.951 & -11.923 & 24.521 & -12.346 & -9.708 \\
J0159.0+3313 & J015905.0+331254 & 15.737 & 0.838 & 14.491 & -12.324 & 24.04 & -12.511 & -9.903 \\
J0209.8+2626 & J020946.5+262528 & 16.719 & 1.143 & 13.209 & -12.906 & 21.518 & -12.123 & -9.795 \\
J0231.0+3505 & J023112.4+350446 & 16.664 & 0.933 & 16.853 & -12.721 & 24.381 & -12.901 & -10.171 \\
J0240.2-0248 & J024004.6-024504 & 14.822 & 0.649 & 13.81 & -11.782 & 21.938 & -12.247 & -9.597 \\
J0259.0+0552 & J025857.4+055243 & 15.377 & 0.625 & 14.791 & -12.062 & 24.04 & -11.873 & -9.598 \\
J0301.6-5617 & J030115.2-561647 & 14.554 & 1.913 & 15.291 & -12.401 & 23.94 & -12.664 & -9.962 \\
J0302.5+3354 & J030226.7+335447 & 15.73 & 0.813 & 16.653 & -12.605 & 23.78 & -12.675 & -9.99 \\
J0327.6+2620 & J032737.2+262007 & 15.973 & 0.772 & 16.593 & -12.521 & 24.741 & -12.732 & -10.018 \\
J0406.2+0639 & J040607.7+063918 & 15.405 & 1.295 & 14.23 & -12.773 & 22.959 & -12.575 & -10.107 \\
J0409.2+2542 & J040921.6+254440 & 15.734 & 0.889 & 16.513 & -12.129 & 25.101 & -12.467 & -9.803 \\
J0427.8-6704 & J042749.5-670435 & 15.044 & 2.093 & 14.23 & -13.143 & 22.338 & -11.696 & -9.422 \\
J0537.5+0959 & J053745.9+095759 & 15.583 & 0.957 & 15.111 & -12.786 & 23.96 & -12.697 & -10.096 \\
J0539.2-6333 & J054002.9-633216 & 14.192 & 1.023 & 15.592 & -11.928 & 24.361 & -12.686 & -9.758 \\
J0610.8-4911 & J061031.8-491222 & 14.529 & 0.557 & 15.792 & -9.862 & 23.98 & -12.752 & -8.073 \\
J0610.8-4911 & J061100.0-491034 & 16.655 & 1.273 & 14.35 & -12.874 & 23.84 & -12.861 & -10.134 \\
J0620.7-5034 & J062045.7-503349 & 15.348 & 0.615 & 14.391 & -12.439 & 23.56 & -12.756 & -9.939 \\
J0633.9+5840 & J063400.1+584035 & 14.913 & 1.062 & 13.75 & -13.233 & 23.72 & -12.796 & -10.34 \\
J0650.6+2055 & J065035.4+205556 & 16.411 & 1.055 & 15.772 & -11.331 & 24.741 & -11.912 & -9.148 \\
J0704.3-4829 & J070421.8-482645 & 16.177 & 0.871 & 16.693 & -12.376 & 24.461 & -12.518 & -9.914 \\
J0738.6+1311 & J073843.4+131330 & 14.496 & 0.517 & 14.491 & -10.78 & 22.499 & -12.031 & -8.916 \\
J0800.1-5531 & J075949.3-553253 & 13.891 & 1.17 & 14.591 & -11.896 & 22.739 & -12.353 & -9.694 \\
J0800.1-5531 & J080013.1-553407 & 13.636 & 1.808 & 14.371 & -12.287 & 22.439 & -12.235 & -9.934 \\
J0800.9+0733 & J080056.5+073235 & 15.365 & 0.779 & 16.653 & -12.037 & 24.701 & -13.152 & -9.849 \\
J0838.5+4013 & J083902.8+401548 & 15.142 & 0.782 & 14.491 & -12.593 & 23.239 & -13.089 & -10.111 \\
J0903.5+4057 & J090342.8+405502 & 14.977 & 0.846 & 16.373 & -12.683 & 24.0 & -12.97 & -10.228 \\
J0906.1-1011 & J090616.2-101430 & 14.9 & 1.836 & 15.552 & -12.787 & 22.999 & -12.516 & -9.981 \\
J0910.1-1816 & J091003.9-181613 & 15.853 & 0.986 & 14.571 & -12.69 & 23.039 & -12.69 & -10.113 \\
J0914.5+6845 & J091429.10+684509 & 15.799 & 0.923 & 16.493 & -12.321 & 24.601 & -12.624 & -9.91 \\
J0928.4-5256 & J092818.7-525700 & 15.527 & 1.159 & 14.21 & -12.696 & 23.099 & -12.021 & -9.696 \\
J0930.9-3030 & J093058.0-303118 & 15.79 & 0.774 & 14.811 & -12.609 & 24.14 & -12.527 & -10.051 \\
J0934.5+7223 & J093334.0+722101 & 13.994 & 1.475 & 14.31 & -12.127 & 22.238 & -12.077 & -9.744 \\
J0938.8+5155 & J093835.0+515455 & 13.151 & 1.428 & 14.511 & -12.712 & 22.459 & -12.655 & -10.102 \\
J1008.2-1000 & J100749.3-094910 & 13.157 & 1.224 & 13.73 & -12.439 & 21.518 & -11.767 & -9.632 \\
J1008.2-1000 & J100802.5-095918 & 12.278 & 0.672 & 16.152 & -12.596 & 23.239 & -13.251 & -10.128 \\
J1008.2-1000 & J100848.6-095450 & 13.842 & 0.921 & 14.451 & -10.533 & 22.459 & -11.726 & -8.724 \\
J1011.1-4420 & J101132.0-442254 & 17.268 & 0.773 & 17.273 & -11.651 & 25.822 & -12.627 & -9.484 \\
J1016.1-4247 & J101620.8-424723 & 15.577 & 0.617 & 15.972 & -11.587 & 24.701 & -12.075 & -9.519 \\
J1018.1-2705 & J101750.2-270550 & 15.534 & 1.811 & 14.13 & -12.655 & 22.118 & -12.023 & -9.672 \\
J1018.1-4051 & J101801.5-405520 & 13.584 & 1.911 & 16.352 & -12.594 & 23.36 & -12.798 & -10.061 \\
J1018.1-4051 & J101807.6-404407 & 13.452 & 2.088 & 12.949 & -12.732 & 23.179 & -12.719 & -10.106 \\
J1024.5-4543 & J102432.6-454428 & 16.675 & 0.985 & 16.993 & -12.767 & 24.12 & -12.494 & -9.985 \\
J1034.7-4645 & J103438.7-464404 & 16.472 & 0.91 & 13.67 & -12.701 & 22.759 & -12.537 & -10.088 \\
J1048.4-5030 & J104824.2-502941 & 15.068 & 0.656 & 16.433 & -11.817 & 25.021 & -12.53 & -9.679 \\
J1049.8+2741 & J104938.8+274217 & 15.567 & 0.756 & 16.593 & -12.44 & 24.2 & -12.701 & -10.005 \\
J1106.7+3623 & J110636.7+362648 & 12.819 & 1.301 & 14.831 & -12.732 & 22.979 & -12.235 & -9.964 \\
J1111.4+0137 & J111114.2+013430 & 16.401 & 1.622 & 14.15 & -12.606 & 23.039 & -12.646 & -10.179 \\
J1119.9-1007 & J111948.2-100704 & 15.428 & 0.701 & 15.031 & -12.428 & 23.42 & -12.434 & -9.85 \\
J1122.0-0231 & J112213.8-022916 & 15.625 & 0.82 & 14.831 & -12.156 & 23.259 & -12.408 & -9.832 \\
J1146.0-0638 & J114600.8-063850 & 16.06 & 0.875 & 16.292 & -12.033 & 25.201 & -12.195 & -9.731 \\
J1155.2-1111 & J115514.7-111125 & 15.624 & 0.728 & 13.69 & -12.906 & 23.58 & -12.936 & -10.284 \\
J1220.1-2458 & J122014.5-245949 & 16.661 & 0.976 & 17.073 & -12.416 & 24.3 & -12.433 & -9.87 \\
J1243.7+1727 & J124351.6+172643 & 15.236 & 0.622 & 14.611 & -12.544 & 23.66 & -12.801 & -10.126 \\
J1256.8+5329 & J125630.4+533203 & 13.797 & 2.319 & 14.03 & -13.105 & 22.058 & -12.034 & -9.816 \\
J1326.0+3507 & J132544.4+350450 & 14.81 & 0.999 & 14.751 & -12.474 & 23.279 & -12.305 & -9.902 \\
J1326.0+3507 & J132622.3+350627 & 14.219 & 2.325 & 15.251 & -12.473 & 23.8 & -12.319 & -9.989 \\
J1415.9-1504 & J141546.1-150228 & 15.109 & 1.199 & 14.35 & -12.315 & 23.099 & -12.478 & -9.94 \\
J1429.8-0739 & J142949.7-073302 & 14.268 & 1.069 & 13.349 & -12.446 & 23.139 & -12.708 & -9.931 \\
J1513.0-3118 & J151244.8-311648 & 14.417 & 0.588 & 15.251 & -10.565 & 23.44 & -12.269 & -8.635 \\
J1514.8+4448 & J151450.10+444957 & 14.531 & 1.852 & 14.711 & -12.316 & 23.119 & -11.8 & -9.728 \\
J1528.4+2004 & J152835.10+200423 & 16.477 & 1.002 & 17.073 & -12.868 & 24.461 & -12.92 & -10.319 \\
J1545.0-6642 & J154459.0-664147 & 16.923 & 0.941 & 17.173 & -11.756 & 25.442 & -12.084 & -9.493 \\
J1557.2+3822 & J155711.9+382032 & 15.668 & 0.778 & 14.21 & -13.033 & 23.239 & -13.012 & -10.488 \\
J1631.8+4144 & J163146.7+414634 & 15.875 & 0.867 & 16.713 & -12.233 & 25.221 & -12.648 & -9.961 \\
J1637.5+3005 & J163728.2+300957 & 13.168 & 1.453 & 13.61 & -12.884 & 22.839 & -13.03 & -10.308 \\
J1637.5+3005 & J163739.3+301015 & 13.39 & 0.968 & 13.289 & -12.06 & 20.937 & -11.607 & -9.482 \\
J1644.8+1850 & J164457.3+185149 & 14.713 & 1.026 & 15.932 & -12.152 & 23.74 & -12.632 & -9.843 \\
J1645.0+1654 & J164500.0+165510 & 15.751 & 0.793 & 14.531 & -12.524 & 23.44 & -12.632 & -10.031 \\
J1651.7-7241 & J165151.5-724309 & 14.62 & 1.34 & 15.151 & -12.814 & 23.079 & -12.846 & -10.141 \\
J1720.6-5144 & J172032.7-514413 & 16.751 & 0.938 & 14.01 & -12.34 & 24.12 & -12.443 & -9.766 \\
J1818.5+2533 & J181830.9+253707 & 14.042 & 0.809 & 14.671 & -11.372 & 22.358 & -11.864 & -9.327 \\
J1846.9-0227 & J184650.7-022903 & 14.29 & 0.455 & 14.631 & -10.252 & 22.959 & -11.081 & -8.262 \\
J1910.8+2856 & J191052.2+285624 & 16.243 & 0.938 & 13.61 & -12.646 & 23.9 & -12.356 & -9.751 \\
J1910.8+2856 & J191059.4+285635 & 15.686 & 0.888 & 14.491 & -12.703 & 24.601 & -12.261 & -9.941 \\
J1918.0+0331 & J191803.6+033030 & 15.64 & 0.88 & 17.113 & -12.641 & 25.281 & -12.199 & -9.867 \\
J1927.5+0154 & J192731.3+015356 & 15.859 & 0.729 & 16.813 & -12.155 & 24.781 & -12.454 & -9.782 \\
J1955.3-5032 & J195512.5-503011 & 15.885 & 1.212 & 14.471 & -12.678 & 22.178 & -12.267 & -9.875 \\
J2008.4+1619 & J200827.6+161843 & 15.952 & 1.637 & 14.27 & -12.715 & 23.339 & -12.513 & -9.995 \\
J2041.1-6138 & J204111.10-613952 & 15.262 & 0.732 & 15.712 & -12.139 & 24.24 & -12.433 & -9.698 \\
J2046.9-5409 & J204700.5-541246 & 15.239 & 1.461 & 15.131 & -12.586 & 24.28 & -12.681 & -10.058 \\
J2109.6+3954 & J210936.4+395513 & 15.81 & 0.744 & 14.19 & -12.945 & 24.32 & -12.879 & -10.276 \\
J2114.9-3326 & J211452.0-332532 & 16.192 & 1.054 & 16.913 & -12.011 & 24.621 & -12.372 & -9.648 \\
J2159.6-4620 & J215935.9-461954 & 15.54 & 0.758 & 16.393 & -12.285 & 24.421 & -12.354 & -9.819 \\
J2207.1+2222 & J220704.3+222234 & 15.346 & 0.627 & 16.473 & -12.51 & 24.481 & -12.621 & -10.024 \\
J2222.9+1507 & J222253.9+151052 & 14.957 & 0.89 & 16.172 & -12.133 & 23.38 & -12.823 & -9.749 \\
J2225.8-0804 & J222552.9-080415 & 15.472 & 1.036 & 15.512 & -12.651 & 24.04 & -12.566 & -10.077 \\
J2237.2-6726 & J223709.3-672614 & 16.106 & 0.784 & 16.773 & -12.537 & 24.22 & -12.62 & -10.019 \\
J2240.3-5241 & J224017.5-524117 & 15.002 & 0.73 & 15.512 & -11.957 & 24.2 & -12.129 & -9.548 \\
J2247.7-5857 & J224744.10-585500 & 15.508 & 1.453 & 16.413 & -12.92 & 23.139 & -12.642 & -10.082 \\
J2303.9+5554 & J230351.7+555618 & 16.605 & 0.8 & 17.353 & -12.483 & 24.781 & -12.558 & -9.951 \\
J2311.6-4427 & J231145.6-443220 & 15.104 & 1.693 & 14.711 & -12.448 & 23.279 & -12.618 & -10.161 \\
J2317.7+2839 & J231740.0+283954 & 15.152 & 0.556 & 16.373 & -11.613 & 25.502 & -12.201 & -9.519 \\
J2326.9-4130 & J232653.2-412713 & 13.578 & 1.096 & 18.294 & -13.669 & 22.819 & -12.335 & -9.453 \\
J2336.9-8427 & J233627.1-842648 & 16.124 & 0.918 & 17.173 & -12.867 & 23.84 & -12.607 & -10.061 \\
J2337.7-2903 & J233730.2-290240 & 15.841 & 0.768 & 16.673 & -12.22 & 23.98 & -12.747 & -9.871 \\
J2351.4-2818 & J235136.5-282154 & 14.196 & 0.587 & 14.811 & -11.302 & 22.899 & -12.485 & -9.208 \\
\enddata
\end{deluxetable*}
\end{longrotatetable}
\begin{longrotatetable}
\begin{deluxetable*}{ccccccccccccc}% C for mathmode
\tablecaption{Synchrotron-self-Compton fits via agnpy with fixed $z = 0.34$ \label{tab:agnpyfits}}
\tablewidth{\columnwidth}
\tablehead{
\colhead{4FGL} & \colhead{XRT} & \colhead{$\log(k_e)$} & \colhead{$p_1$} & \colhead{$p_2$} & \colhead{$\log(\gamma_b)$} & \colhead{$\log(\gamma_{max})$} & \colhead{$\log(\gamma_{min})$} & \colhead{$\delta_D$} & \colhead{$\log(B)$} & \colhead{$\log(t_{var})$} & \colhead{$\chi^2_r$} \\
\colhead{} & \colhead{} & \colhead{} & \colhead{} & \colhead{} & \colhead{} & \colhead{} & \colhead{} & \colhead{} & \colhead{$\rm{G}$} & \colhead{$\rm{s}$} & \colhead{}
}
\startdata
J0004.4-4001 & J000434.1-400036 & -6.652 & 2.19 & 4.6 & 4.402 & 5.485 & 2.17 & 16.323 & -1.151 & 5.309 & 0.715 \\
J0025.4-4838 & J002536.8-483808 & -5.0 & 4.0 & 3.731 & 4.298 & 7.677 & 3.493 & 10.579 & -0.382 & 4.377 & 0.805 \\
J0026.1-0732 & J002611.9-073115 & -4.848 & 2.06 & 3.453 & 4.094 & 6.061 & 1.945 & 10.928 & -0.803 & 4.925 & 1.469 \\
J0031.5-5648 & J003135.1-564640 & -1.621 & -1.002 & 2.738 & 2.983 & 5.407 & 3.38 & 24.931 & -1.241 & 4.196 & 0.581 \\
J0037.2-2653 & J003729.6-265043 & -1.965 & -0.931 & 3.383 & 3.165 & 5.661 & 2.903 & 23.844 & -1.496 & 4.57 & 0.648 \\
J0057.9+6326 & J005758.1+632642 & -0.371 & -1.195 & 3.317 & 2.847 & 7.059 & 2.861 & 11.405 & -0.976 & 4.728 & 2.952 \\
J0058.3-4603 & J005806.4-460417 & -5.625 & 3.451 & 2.45 & 4.433 & 5.742 & 3.368 & 8.202 & -0.521 & 4.646 & 0.499 \\
J0118.3-6008 & J011823.10-600753 & -4.495 & 3.372 & 3.704 & 4.001 & 5.652 & 3.743 & 13.172 & -1.21 & 5.042 & 1.048 \\
J0120.2-7944 & J011914.7-794509 & -4.808 & 3.996 & 3.482 & 4.254 & 5.741 & 3.613 & 12.8 & -0.743 & 4.352 & 0.994 \\
J0125.9-6303 & J012548.3-630244 & -4.863 & 3.999 & 2.953 & 3.94 & 5.663 & 3.693 & 17.898 & -1.352 & 4.851 & 0.791 \\
J0156.3-2420 & J015624.4-242004 & -4.826 & -1.287 & 3.453 & 3.997 & 5.66 & 3.983 & 22.157 & -1.395 & 4.889 & 1.503 \\
J0159.0+3313 & J015905.0+331254 & -4.958 & 2.907 & 3.237 & 4.046 & 5.623 & 3.61 & 13.411 & -1.082 & 5.006 & 2.018 \\
J0209.8+2626 & J020946.5+262528 & -2.653 & 1.552 & 3.419 & 3.633 & 5.017 & 2.626 & 14.686 & -1.955 & 4.971 & 1.001 \\
J0231.0+3505 & J023112.4+350446 & -3.452 & 1.707 & 2.968 & 3.443 & 5.997 & 3.54 & 12.505 & -1.248 & 5.127 & 1.857 \\
J0240.2-0248 & J024004.6-024504 & -6.936 & 3.707 & 3.875 & 4.443 & 6.084 & 3.463 & 21.781 & -1.422 & 5.024 & 1.218 \\
J0259.0+0552 & J025857.4+055243 & -3.798 & 2.439 & 3.645 & 3.927 & 5.822 & 3.975 & 13.722 & -1.303 & 5.027 & 1.848 \\
J0301.6-5617 & J030115.2-561647 & -6.674 & 2.476 & 4.047 & 4.607 & 6.218 & 2.136 & 20.002 & -1.524 & 4.911 & 0.93 \\
J0302.5+3354 & J030226.7+335447 & -4.213 & 2.438 & 2.869 & 3.72 & 5.665 & 2.331 & 15.674 & -1.382 & 4.962 & 1.119 \\
J0327.6+2620 & J032737.2+262007 & -6.034 & 2.23 & 2.896 & 4.235 & 5.863 & 2.406 & 16.458 & -1.688 & 5.264 & 0.81 \\
J0406.2+0639 & J040607.7+063918 & -3.654 & 1.7 & 3.318 & 3.757 & 5.662 & 1.034 & 6.628 & -0.864 & 5.297 & 1.219 \\
J0409.2+2542 & J040921.6+254440 & -8.412 & 2.401 & 6.258 & 5.307 & 6.283 & 2.352 & 22.337 & -1.56 & 4.815 & 0.762 \\
J0427.8-6704 & J042749.5-670435 & -3.284 & 2.509 & 4.659 & 4.34 & 7.908 & 3.008 & 22.282 & -1.979 & 4.0 & 1.647 \\
J0537.5+0959 & J053745.9+095759 & -5.297 & 2.529 & 3.117 & 4.246 & 5.645 & 2.365 & 12.992 & -1.252 & 4.911 & 2.84 \\
J0539.2-6333 & J054002.9-633216 & -7.435 & 2.357 & 5.257 & 4.687 & 7.341 & 3.11 & 18.911 & -1.245 & 5.131 & 0.987 \\
J0610.8-4911 & J061031.8-491222 & -7.499 & -0.87 & 7.997 & 3.996 & 5.399 & 3.634 & 31.275 & -0.189 & 5.427 & 1.391 \\
J0610.8-4911 & J061100.0-491034 & -5.528 & 3.11 & 3.057 & 4.486 & 7.949 & 3.296 & 11.25 & -0.501 & 4.326 & 0.526 \\
J0620.7-5034 & J062045.7-503349 & -6.098 & 2.392 & 3.208 & 4.168 & 5.824 & 2.366 & 23.116 & -1.997 & 5.29 & 1.155 \\
J0633.9+5840 & J063400.1+584035 & -4.26 & 1.634 & 3.176 & 3.874 & 6.027 & 2.417 & 9.387 & -1.842 & 5.518 & 1.668 \\
J0650.6+2055 & J065035.4+205556 & -6.215 & 1.406 & 3.426 & 4.368 & 5.501 & 1.191 & 18.074 & -0.957 & 5.074 & 0.345 \\
J0704.3-4829 & J070421.8-482645 & -4.976 & 2.966 & 2.866 & 4.088 & 5.503 & 3.554 & 13.8 & -0.949 & 4.788 & 0.371 \\
J0738.6+1311 & J073843.4+131330 & -5.288 & 3.626 & 4.994 & 3.796 & 6.016 & 3.619 & 27.845 & -1.064 & 5.195 & 2.574 \\
J0800.1-5531 & J075949.3-553253 & -5.13 & 1.823 & 4.307 & 4.008 & 7.438 & 2.406 & 10.272 & -0.88 & 5.433 & 0.546 \\
J0800.1-5531 & J080013.1-553407 & -3.446 & 3.999 & 3.994 & 3.85 & 6.603 & 3.546 & 11.35 & -0.7 & 4.616 & 0.98 \\
J0800.9+0733 & J080056.5+073235 & -3.343 & 1.596 & 2.511 & 2.749 & 5.461 & 2.478 & 14.0 & -1.23 & 5.537 & 2.624 \\
J0838.5+4013 & J083902.8+401548 & -5.27 & 2.525 & 3.14 & 3.932 & 5.583 & 2.238 & 17.695 & -1.348 & 5.032 & 1.662 \\
J0903.5+4057 & J090342.8+405502 & -4.187 & 2.162 & 2.792 & 3.631 & 5.461 & 2.416 & 12.706 & -1.217 & 5.091 & 0.633 \\
J0906.1-1011 & J090616.2-101430 & -6.327 & 2.942 & 3.476 & 4.74 & 5.42 & 2.808 & 13.53 & -0.858 & 4.454 & 0.525 \\
J0910.1-1816 & J091003.9-181613 & -3.586 & 1.934 & 3.166 & 3.662 & 5.474 & 1.872 & 4.113 & -0.206 & 5.431 & 1.512 \\
J0914.5+6845 & J091429.10+684509 & -8.271 & 2.549 & 6.052 & 5.26 & 7.978 & 2.189 & 19.984 & -1.373 & 4.728 & 0.868 \\
J0928.4-5256 & J092818.7-525700 & -3.339 & 2.918 & 3.417 & 4.01 & 5.963 & 3.388 & 12.625 & -0.995 & 4.382 & 1.196 \\
J0930.9-3030 & J093058.0-303118 & -4.275 & 1.466 & 3.227 & 4.036 & 5.655 & 1.015 & 23.599 & -1.48 & 4.398 & 0.688 \\
J0934.5+7223 & J093334.0+722101 & -4.41 & 1.826 & 5.781 & 4.019 & 7.194 & 2.513 & 8.466 & -0.968 & 5.38 & 0.739 \\
J0938.8+5155 & J093835.0+515455 & -5.437 & 2.839 & 3.725 & 4.268 & 5.67 & 2.986 & 10.506 & -1.07 & 5.062 & 0.912 \\
J1008.2-1000 & J100749.3-094910 & -4.962 & 4.0 & 7.004 & 4.427 & 6.438 & 3.376 & 20.832 & -1.281 & 4.021 & 1.874 \\
J1008.2-1000 & J100802.5-095918 & -4.843 & 2.725 & 2.847 & 3.665 & 5.353 & 1.82 & 17.248 & -1.339 & 5.105 & 3.84 \\
J1008.2-1000 & J100848.6-095450 & -5.284 & 1.604 & 6.832 & 3.82 & 7.237 & 3.687 & 27.338 & -1.157 & 5.393 & 0.896 \\
J1011.1-4420 & J101132.0-442254 & -6.402 & 3.016 & 2.26 & 3.773 & 6.028 & 1.82 & 14.274 & -1.94 & 6.015 & 0.966 \\
J1016.1-4247 & J101620.8-424723 & -6.032 & 2.157 & 5.036 & 4.565 & 5.472 & 3.937 & 19.153 & -0.718 & 4.542 & 2.913 \\
J1018.1-2705 & J101750.2-270550 & -5.168 & 2.491 & 6.696 & 4.418 & 5.52 & 2.652 & 15.713 & -1.948 & 5.058 & 1.454 \\
J1018.1-4051 & J101801.5-405520 & -1.512 & 1.182 & 2.8 & 2.469 & 5.589 & 2.733 & 7.348 & -1.302 & 5.919 & 2.289 \\
J1018.1-4051 & J101807.6-404407 & -4.582 & 3.5 & 2.999 & 3.852 & 5.564 & 3.072 & 9.331 & -1.165 & 5.279 & 1.443 \\
J1024.5-4543 & J102432.6-454428 & -3.666 & 3.996 & 2.903 & 3.92 & 5.62 & 3.39 & 19.133 & -0.857 & 4.0 & 1.088 \\
J1034.7-4645 & J103438.7-464404 & -4.961 & 3.999 & 3.419 & 4.138 & 7.008 & 3.71 & 18.373 & -1.954 & 4.998 & 1.247 \\
J1048.4-5030 & J104824.2-502941 & -6.609 & 3.969 & 2.101 & 4.197 & 5.342 & 3.732 & 19.891 & -1.533 & 5.234 & 1.695 \\
J1049.8+2741 & J104938.8+274217 & -3.408 & 2.054 & 2.673 & 3.39 & 5.453 & 2.146 & 16.0 & -1.237 & 4.85 & 2.156 \\
J1106.7+3623 & J110636.7+362648 & -3.553 & 2.085 & 6.715 & 4.226 & 7.973 & 1.042 & 17.753 & -1.071 & 4.0 & 0.653 \\
J1111.4+0137 & J111114.2+013430 & -4.444 & 3.997 & 3.518 & 3.979 & 7.116 & 3.7 & 12.205 & -1.305 & 5.045 & 2.075 \\
J1119.9-1007 & J111948.2-100704 & -5.732 & 2.518 & 3.82 & 4.419 & 5.71 & 2.409 & 18.917 & -1.405 & 4.748 & 0.426 \\
J1122.0-0231 & J112213.8-022916 & -6.188 & 2.247 & 5.838 & 4.44 & 6.748 & 2.186 & 15.023 & -1.415 & 5.242 & 1.175 \\
J1146.0-0638 & J114600.8-063850 & -3.927 & 3.765 & 2.037 & 3.404 & 5.221 & 2.756 & 16.351 & -1.369 & 4.915 & 2.015 \\
J1155.2-1111 & J115514.7-111125 & -3.938 & 1.495 & 3.155 & 3.65 & 5.962 & 2.12 & 14.884 & -1.671 & 5.162 & 1.249 \\
J1220.1-2458 & J122014.5-245949 & -4.456 & 1.624 & 2.836 & 3.863 & 5.817 & 1.026 & 8.959 & -1.008 & 5.266 & 0.384 \\
J1243.7+1727 & J124351.6+172643 & -4.532 & 1.708 & 3.215 & 3.893 & 5.453 & 1.774 & 14.445 & -1.153 & 4.926 & 1.112 \\
J1256.8+5329 & J125630.4+533203 & -3.302 & 2.326 & 5.733 & 4.234 & 5.01 & 3.004 & 25.007 & -1.996 & 4.037 & 1.448 \\
J1326.0+3507 & J132544.4+350450 & -3.811 & 1.674 & 3.41 & 3.876 & 5.367 & 1.826 & 5.777 & -0.476 & 5.217 & 0.998 \\
J1326.0+3507 & J132622.3+350627 & -4.469 & 1.945 & 4.811 & 4.284 & 6.436 & 1.07 & 10.137 & -0.68 & 4.63 & 0.283 \\
J1415.9-1504 & J141546.1-150228 & -5.12 & 3.136 & 3.559 & 4.154 & 5.637 & 3.497 & 11.266 & -0.859 & 4.965 & 0.847 \\
J1429.8-0739 & J142949.7-073302 & -3.081 & 3.999 & 3.061 & 3.313 & 5.503 & 3.031 & 8.524 & -0.646 & 5.218 & 3.172 \\
J1513.0-3118 & J151244.8-311648 & -6.8 & 1.325 & 5.507 & 4.069 & 7.108 & 3.026 & 17.988 & -0.553 & 5.619 & 1.963 \\
J1514.8+4448 & J151450.10+444957 & -4.98 & 3.885 & 7.993 & 4.448 & 7.649 & 3.738 & 9.709 & -0.692 & 4.56 & 1.512 \\
J1528.4+2004 & J152835.10+200423 & -3.493 & 1.668 & 2.786 & 3.606 & 5.78 & 1.66 & 13.503 & -1.183 & 4.712 & 0.99 \\
J1545.0-6642 & J154459.0-664147 & -8.895 & 2.311 & 6.633 & 5.589 & 7.189 & 3.348 & 16.974 & -1.375 & 4.98 & 1.635 \\
J1557.2+3822 & J155711.9+382032 & -4.231 & 1.42 & 3.349 & 3.868 & 5.949 & 1.154 & 8.143 & -1.209 & 5.317 & 0.892 \\
J1631.8+4144 & J163146.7+414634 & -0.263 & -0.574 & 2.384 & 2.0 & 5.422 & 2.708 & 22.615 & -1.358 & 4.546 & 0.917 \\
J1637.5+3005 & J163728.2+300957 & -4.399 & 2.4 & 3.218 & 3.732 & 5.821 & 2.813 & 15.826 & -1.741 & 5.226 & 0.805 \\
J1637.5+3005 & J163739.3+301015 & -2.613 & 2.326 & 4.503 & 3.58 & 6.627 & 3.41 & 13.662 & -1.562 & 5.144 & 0.588 \\
J1644.8+1850 & J164457.3+185149 & -5.648 & 2.594 & 4.877 & 4.482 & 5.242 & 2.789 & 18.771 & -0.394 & 4.091 & 1.494 \\
J1645.0+1654 & J164500.0+165510 & -4.693 & 2.03 & 3.486 & 4.106 & 5.898 & 2.192 & 41.146 & -1.902 & 4.225 & 1.035 \\
J1651.7-7241 & J165151.5-724309 & -6.914 & 2.801 & 3.412 & 4.64 & 5.832 & 2.541 & 16.39 & -1.598 & 5.047 & 1.446 \\
J1720.6-5144 & J172032.7-514413 & -4.116 & 3.983 & 3.053 & 3.75 & 6.059 & 3.453 & 14.546 & -1.113 & 4.932 & 1.393 \\
J1818.5+2533 & J181830.9+253707 & -3.025 & -0.419 & 4.725 & 3.624 & 7.825 & 1.489 & 21.021 & -0.516 & 4.362 & 0.592 \\
J1846.9-0227 & J184650.7-022903 & -6.155 & 0.669 & 4.816 & 4.021 & 5.424 & 1.114 & 18.431 & -1.222 & 5.976 & 2.013 \\
J1910.8+2856 & J191052.2+285624 & -3.958 & 1.002 & 3.096 & 3.678 & 7.808 & 2.402 & 17.61 & -1.999 & 5.27 & 1.542 \\
J1910.8+2856 & J191059.4+285635 & -2.793 & 2.647 & 3.124 & 3.545 & 5.876 & 3.999 & 20.222 & -1.971 & 4.862 & 1.225 \\
J1918.0+0331 & J191803.6+033030 & -4.601 & 2.181 & 2.46 & 4.182 & 5.864 & 1.401 & 14.16 & -1.692 & 4.755 & 2.079 \\
J1927.5+0154 & J192731.3+015356 & -1.884 & -1.766 & 2.513 & 2.668 & 5.572 & 2.491 & 15.004 & -1.333 & 5.06 & 1.404 \\
J1955.3-5032 & J195512.5-503011 & -4.586 & 2.697 & 4.135 & 4.178 & 7.901 & 2.898 & 7.503 & -0.871 & 5.101 & 1.724 \\
J2008.4+1619 & J200827.6+161843 & -2.876 & 1.72 & 3.207 & 3.649 & 5.695 & 1.606 & 18.052 & -1.117 & 4.31 & 0.787 \\
J2041.1-6138 & J204111.10-613952 & -8.348 & 2.548 & 3.729 & 4.963 & 5.729 & 2.342 & 14.618 & -1.717 & 5.639 & 1.263 \\
J2046.9-5409 & J204700.5-541246 & -6.715 & 2.375 & 3.409 & 4.567 & 5.887 & 1.868 & 16.373 & -1.769 & 5.256 & 0.725 \\
J2109.6+3954 & J210936.4+395513 & -3.971 & 2.289 & 3.029 & 3.701 & 5.624 & 2.317 & 20.009 & -1.454 & 4.659 & 3.829 \\
J2114.9-3326 & J211452.0-332532 & -2.212 & 3.853 & 2.543 & 2.737 & 5.622 & 2.605 & 12.414 & -1.218 & 5.293 & 1.088 \\
J2159.6-4620 & J215935.9-461954 & -4.619 & 1.558 & 2.859 & 3.893 & 5.502 & 3.421 & 12.142 & -1.137 & 5.141 & 1.038 \\
J2207.1+2222 & J220704.3+222234 & -4.782 & 2.209 & 2.711 & 3.938 & 5.527 & 2.304 & 18.931 & -1.544 & 4.838 & 0.79 \\
J2222.9+1507 & J222253.9+151052 & -5.513 & 2.6 & 2.837 & 3.862 & 5.223 & 2.314 & 10.737 & -0.87 & 5.47 & 0.431 \\
J2225.8-0804 & J222552.9-080415 & -5.403 & 2.368 & 4.134 & 4.506 & 7.954 & 1.917 & 12.606 & -0.924 & 4.577 & 1.148 \\
J2237.2-6726 & J223709.3-672614 & -4.334 & 2.063 & 2.795 & 3.825 & 5.544 & 1.797 & 8.423 & -0.865 & 5.183 & 0.572 \\
J2240.3-5241 & J224017.5-524117 & -6.465 & 2.348 & 4.262 & 4.598 & 5.399 & 2.183 & 15.088 & -1.169 & 5.055 & 0.757 \\
J2247.7-5857 & J224744.10-585500 & -2.493 & 2.02 & 2.953 & 3.228 & 5.703 & 2.568 & 7.819 & -1.047 & 5.29 & 1.75 \\
J2303.9+5554 & J230351.7+555618 & -3.744 & 2.701 & 2.64 & 3.351 & 6.187 & 2.924 & 15.126 & -2.0 & 5.437 & 2.605 \\
J2311.6-4427 & J231145.6-443220 & -4.441 & 2.21 & 3.819 & 4.023 & 5.553 & 3.6 & 10.379 & -0.815 & 4.93 & 0.82 \\
J2317.7+2839 & J231740.0+283954 & -6.031 & 3.165 & 1.377 & 4.071 & 5.1 & 2.768 & 21.651 & -1.284 & 4.82 & 0.725 \\
J2326.9-4130 & J232653.2-412713 & -4.842 & 2.334 & 4.902 & 3.998 & 7.708 & 2.485 & 7.891 & 0.158 & 5.0 & 0.608 \\
J2336.9-8427 & J233627.1-842648 & -6.009 & 3.76 & 2.174 & 4.586 & 5.317 & 3.478 & 10.276 & -0.48 & 4.391 & 0.446 \\
J2337.7-2903 & J233730.2-290240 & -0.815 & 2.583 & 2.611 & 2.375 & 5.141 & 2.087 & 8.092 & -0.346 & 5.041 & 0.696 \\
J2351.4-2818 & J235136.5-282154 & -6.182 & 2.059 & 4.489 & 4.059 & 5.468 & 2.12 & 21.662 & -1.009 & 5.258 & 1.079 \\
\enddata
\end{deluxetable*}
\end{longrotatetable}

\end{document}